\documentclass[aps,superscriptaddress,twocolumn]{revtex4}
\usepackage{mathrsfs}
\usepackage{amsfonts}
\usepackage{amsmath}
\usepackage{amssymb}
\usepackage{graphicx}

\providecommand{\func}[1]{\operatorname{#1}}
\providecommand{\U}[1]{\protect\rule{.1in}{.1in}}

\begin{document}

\title{Resonant-impurity scanning tunneling spectroscopy in altermagnets:
dual Fano resonance and Landau-quantization-induced nodal spin contrast}
\author{Yuan Hong}
\affiliation{National Key Laboratory of Computational Physics, Institute of Applied
Physics and Computational Mathematics, Beijing 100088, China}
\affiliation{Graduate School, China Academy of Engineering Physics, Beijing 100193, China}
\author{Zhigang Wang}
\affiliation{National Key Laboratory of Computational Physics, Institute of Applied
Physics and Computational Mathematics, Beijing 100088, China}
\author{Zhen-Guo Fu}
\thanks{Corresponding author. Email address: fu\_zhenguo@iapcm.ac.cn}
\affiliation{National Key Laboratory of Computational Physics, Institute of Applied
Physics and Computational Mathematics, Beijing 100088, China}
\author{Feng Chi}
\affiliation{School of Electronic and Information Engineering, Zhongshan Institute,
University of Electronic Science and Technology of China, Zhongshan 528400,
China}
\author{Cong Wang}
\affiliation{National Key Laboratory of Computational Physics, Institute of Applied
Physics and Computational Mathematics, Beijing 100088, China}
\author{Wei Zhang}
\thanks{Corresponding author. Email address: zhang\_wei@iapcm.ac.cn}
\affiliation{National Key Laboratory of Computational Physics, Institute of Applied
Physics and Computational Mathematics, Beijing 100088, China}
\author{Ping Zhang}
\thanks{Corresponding author. Email address: zhang\_ping@iapcm.ac.cn}
\affiliation{National Key Laboratory of Computational Physics, Institute of Applied
Physics and Computational Mathematics, Beijing 100088, China}
\affiliation{School of Physics and Physical Engineering, Qufu Normal University, Qufu
273165, China}

\begin{abstract}
Using a Green's-function formalism, we study the spin-resolved local
spectral function of a resonant impurity coupled to a two-dimensional $d$%
-wave altermagnetic substrate. It is found that the interplay between direct
tunneling from the impurity to the scanning tunneling microscopy (STM) tip
and altermagnet-mediated tunneling gives rise to a dual Fano resonance in
the absence of an external magnetic field. Moreover, the anisotropic
spin-dependent oscillations of the local density of states and the
corresponding Fano factors provide information on the altermagnetic
splitting strength from complementary local and global perspectives. In
addition, spin-selective tunneling can be achieved by tuning the Fermi
energy and the tip position. In the presence of a strong magnetic field with
Landau-level quantization, the dominant scanning tunneling spectroscopy
(STS) signature appears as a spin-dependent nodal structure in real space:
the nodal mismatch between opposite spin channels produces a large local
spin contrast. These results establish resonant-impurity STM/STS as a
phase-sensitive local probe of altermagnetic band anisotropy.
\end{abstract}

\maketitle

\section{Introduction}

Altermagnets form a class of collinear spin-compensated magnets whose
electronic structures are governed by spin-group symmetries with decoupled
real-space crystal operations and spin-space transformations. As a
consequence, they can have vanishing net magnetization, as in
antiferromagnets, while exhibiting sizable momentum-dependent spin splitting
even in the nonrelativistic limit \cite%
{Smejkal2022Emerging,Mazin2022Altermagnetism,Zeng2024TwoDimensionalAltermagnetism,GonzalezHernandez2021SpinSplitter,Smejkal2022GiantTMR}%
. The resulting anisotropic spin-split Fermi surfaces have motivated studies
of spin-dependent transport, spin-current responses, and spintronic effects
in compensated magnetic systems \cite%
{Bai2022SpinSplittingTorque,Karube2022SpinSplitterTorque,GonzalezBetancourt2023AHE,McClarty2024LandauTheory,Roig2024MinimalModels}%
. Candidate materials, including MnTe, CrSb, MnTe$_{2}$, KV$_{2}$Se$_{2}$O,
Mn$_{5}$Si$_{3}$, and Rb$_{1-\delta }$V$_{2}$Te$_{2}$O, provide concrete
platforms for studying altermagnetic electronic structures \cite%
{Krempasky2024Kramers,Lee2024MnTe,Osumi2024MnTe,Reimers2024CrSbThinFilm,Zhu2024Plaid,Jiang2025KV2Se2O,Zhang2025RbV2Te2O,Reichlova2024Mn5Si3AHE,Badura2025Mn5Si3ANE}%
.

Local tunneling probes provide information complementary to momentum-space
spectroscopy and bulk transport. In scanning tunneling microscopy (STM),
impurity-induced modulations of the local density of states are governed by
quasiparticle propagation between the impurity and the tip. These
modulations therefore carry information about the Fermi-surface anisotropy
and the spin-group symmetry of the host material \cite%
{Tersoff1983STM,Tersoff1985STM,Friedel1952,Crommie1993StandingWaves,Hasegawa1993StandingWaves,Sprunger1997Friedel}%
. Recent theoretical work has shown that impurities in altermagnets can
produce spin-resolved Friedel oscillations, quasiparticle-interference
patterns, and characteristic local features \cite%
{Chen2024ImpurityScatteringAltermagnets,Sukhachov2024ImpurityInducedFriedel,Hu2025QuasiparticleInterference,Gondolf2025LocalSignaturesAltermagnetism,Maiani2025ImpurityStates}%
. These results indicate that real-space patterns measured by scanning
tunneling spectroscopy (STS) can directly reflect the spin-dependent band
geometry of altermagnets.

Two key aspects of impurity-induced local STS responses in altermagnets
remain largely unexplored. First, altermagnetic anisotropy may manifest not
only in such real-space patterns, but also in complementary energy-domain
line shapes. Existing studies have mainly addressed the former aspect, for
which modeling the impurity as a scattering potential is often sufficient
because it captures the leading spatial signatures of impurity scattering 
\cite%
{Chen2024ImpurityScatteringAltermagnets,Sukhachov2024ImpurityInducedFriedel,Hu2025QuasiparticleInterference,Gondolf2025LocalSignaturesAltermagnetism}%
. However, to describe the latter spectroscopic response, especially for
resonant impurities, one must go beyond a potential-scattering description
and include an explicit impurity orbital degree of freedom. On the one hand,
impurity-substrate hybridization renormalizes the impurity resonance and
embeds it in the perturbed substrate continuum, giving rise to an
energy-dependent resonant feature in the energy domain. On the other hand,
the impurity orbital can couple directly to the STM tip, opening an
additional tunneling pathway. These two mechanisms can both generate
Fano-type line-shape asymmetries \cite%
{Fano1961,Madhavan1998Kondo,schiller2000theory,Ujsaghy2000FanoSTM,plihal2001nonequilibrium,knorr2002kondo}%
, thereby providing complementary spectroscopic probes of the underlying
spin-dependent anisotropy of the altermagnetic substrate.

Second, the magnetic-field dependence of impurity-induced real-space
patterns has not been systematically analyzed in altermagnets. In
particular, the orbital effect of an out-of-plane magnetic field introduces
Landau quantization, which can greatly modify the real-space structure of
the substrate. Since spin-compensated collinear magnetic structures can
remain stable under relatively strong magnetic fields before field-induced
transitions occur, high-field physics of candidate altermagnetic systems is
of direct interest \cite{Terashima2026CrSbQO,Wu2025RuO2QO}. Resonant
impurity scattering in a magnetic field thus provides a setting in which
impurity resonance, altermagnetic anisotropy, direct tip-impurity coupling,
altermagnet-mediated coupling, and Landau quantization enter the local
spin-resolved tunneling signal on the same footing.

Considering these unclear issues, we study a resonant impurity coupled to a
two-dimensional $d$-wave altermagnetic substrate in a scanning tunneling
geometry, paying attention to the interference among different
(spin-dependent) tunneling channels. Using a Green's-function formalism, we
derive the spin-resolved local spectral function measured by the tip. In the
case of zero magnetic field, we identify a dual Fano resonance in the energy
domain. In the absence of direct tip-impurity coupling (for large
tip-impurity separation), the impurity resonance is encoded in the perturbed
substrate continuum, leading to a Fano-type line shape whose asymmetry is
controlled by the phase of the substrate Green's function connecting the
impurity and the tip. This Fano profile produces spin-mismatched line shapes
and allows the altermagnetic splitting parameter $\mathcal{J}$ to be
extracted from the spatial periods of both the impurity-induced oscillations
and the Fano $q$ factor. In the presence of direct tip-impurity coupling
(for small tip-impurity separation), the mechanism of Fano resonance crosses
over to the other picture, namely the interference between the direct
tip-impurity tunneling path and the substrate-mediated tunneling path. In
this case, spin-selective tunneling can be achieved by properly tuning the
Fermi energy and the tip position. In the Landau-quantized regime with
strong magnetic field, the dominant signature is instead the real-space
pattern: the mismatch between nodal and non-nodal regions of opposite spin
channels produces a large local spin contrast , while impurity detuning and
the Landau-level filling factor provide additional control over the
resonance condition. These results establish resonant-impurity STM/STS as a
phase-sensitive local probe of altermagnetic electronic structure.

\section{Model Hamiltonian and Formalism}

\begin{figure}[ptb]
\centering
\includegraphics[width=1\linewidth]{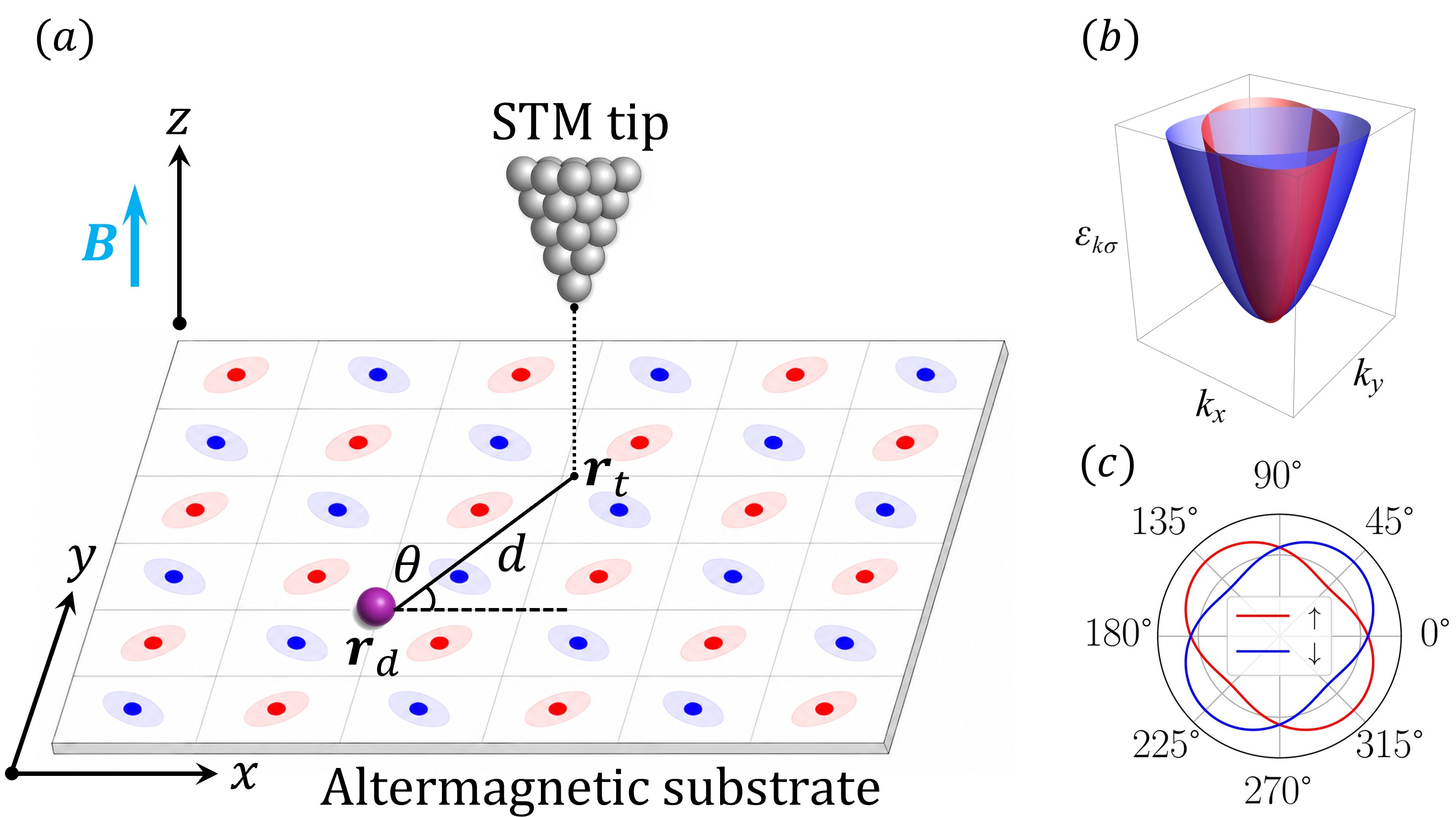}
\caption{(a) Schematic illustration of the STM/STS setup on an altermagnetic
substrate in an out-of-plane magnetic field. The impurity is shown as a
purple sphere, and the red and blue dots denote substrate atoms with
opposite spin moments. The surrounding ellipses indicate anisotropic
spin-dependent charge-density contours related by the $\hat{C}_{4}\hat{T}$
symmetry. (b) Dispersion of the $d$-wave altermagnetic substrate. Red and
blue surfaces represent the two spin branches. (c) Spin-dependent anisotropy
factor $\protect\kappa_{\protect\sigma}(\protect\theta)$ for $\mathcal{J}
=0.4$.}
\label{fig:setup_model}
\end{figure}

Our setup is shown in Fig.~\ref{fig:setup_model}(a). A resonant impurity is
placed on a two-dimensional $d$-wave altermagnetic substrate subject to an
out-of-plane magnetic field. The system is described by a noninteracting
resonant-level model, 
\begin{equation}
\left\{ 
\begin{array}{c}
H=H_{0}+V, \\[5pt] 
H_{0}=H_{s}+H_{d},%
\end{array}
\right.
\end{equation}
The impurity Hamiltonian is 
\begin{equation}
H_{d}=\sum_{\sigma}\mathcal{E}_{\sigma}d_{\sigma}^{\dagger}d_{\sigma},
\end{equation}
where $\mathcal{E}_{\sigma}=\varepsilon_{0}+\sigma\varepsilon_{d}$. Here $%
\varepsilon_{0}$ is the bare impurity level, and $\varepsilon_{d}$ denotes
the impurity Zeeman energy.

The substrate Hamiltonian is 
\begin{equation}
H_{s}=\sum_{\boldsymbol{k},\sigma }\varepsilon _{\boldsymbol{k}\sigma }c_{%
\boldsymbol{k}\sigma }^{\dagger }c_{\boldsymbol{k}\sigma },
\end{equation}%
where $c_{\boldsymbol{k}\sigma }$ and $d_{\sigma }$ are the annihilation
operators of substrate state $\left\vert \boldsymbol{k}\sigma \right\rangle $
and the impurity state $\left\vert d\sigma \right\rangle $, respectively.

In the zero (magnetic)-field regime, the substrate dispersion is written as 
\cite{ezawa2026tunneling} 
\begin{equation}
\varepsilon_{\boldsymbol{k}\sigma}=\frac{\hbar^{2}}{2m}\left( k_{x}^{2}
+k_{y}^{2}+2\sigma\mathcal{J}k_{x}k_{y}\right) ,
\end{equation}
where $\sigma=\pm1$ (or $\uparrow/\downarrow$) labels the spin-up and
spin-down states, and $\mathcal{J}$ is the dimensionless altermagnetic
splitting strength. We focus on $\mathcal{J}<1$, for which the two
spin-split Fermi surfaces are ellipses related by the $\hat{C}_{4}\hat{T}$
symmetry [see Fig.~\ref{fig:setup_model}(b)]. Here $\hat{C}_{4}$ denotes a
fourfold crystal rotation acting only on the spatial sector, while $\hat{T}$
denotes time reversal.

In the Landau-quantized regime, where the orbital magnetic-field effect is
resolved, the single-particle Hamiltonian with minimal coupling is 
\begin{equation}
\hat{h}_{s}=\frac{1}{2m}\left[ \hat{\pi}_{x}^{2}+\hat{\pi}_{y}^{2}+\sigma 
\mathcal{J}\left( \hat{\pi}_{x}\hat{\pi}_{y}+\hat{\pi}_{y}\hat{\pi}%
_{x}\right) \right] ,
\end{equation}%
where $\hat{\pi}_{i}=\hat{p}_{i}+eA_{i}/c$ is the kinetic momentum operator,
and the Landau gauge $\boldsymbol{A}=Bx\hat{\boldsymbol{e}}_{y}$ is used.
Throughout this work, the substrate Zeeman splitting is neglected, while the
impurity Zeeman splitting is retained as a spin-dependent tuning parameter.
Away from Landau-level resonances, this approximation is justified by the
separation between the substrate energy scale and the much narrower impurity
linewidth, as discussed in Appendix~\ref{app:impurity_renormalization}.

The hybridization between the impurity and the substrate is described by 
\begin{equation}
V=\sum_{\boldsymbol{k},\sigma }\left( V_{\boldsymbol{k}\sigma }d_{\sigma
}^{\dagger }c_{\boldsymbol{k}\sigma }+V_{\boldsymbol{k}\sigma }^{\ast }c_{%
\boldsymbol{k}\sigma }^{\dagger }d_{\sigma }\right) .
\end{equation}%
Within the contact-potential approximation, the corresponding tunneling
matrix element is 
\begin{equation}
V_{\boldsymbol{k}\sigma }=\left\langle d\sigma \right\vert \hat{V}\left\vert 
\boldsymbol{k}\sigma \right\rangle \approx \mathcal{A}_{ai}\left\langle 
\boldsymbol{r}_{d}\sigma \middle|\boldsymbol{k}\sigma \right\rangle ,
\label{eq:hyb_matrix_element}
\end{equation}%
where $\boldsymbol{r}_{d}$ is the in-plane projection of the impurity
position, and $\mathcal{A}_{ai}$ characterizes the impurity-substrate
coupling.

The spin-resolved STM signal measured at $\boldsymbol{r}_{t}$ is
proportional to the local spectral function \cite{plihal2001nonequilibrium}, 
\begin{equation}
A_{\sigma}(d,\theta) = -2\func{Im} \left\langle \boldsymbol{r}
_{t}\sigma\left\vert \hat{V}_{t}\hat{G}^{R}(E_{F})\hat{V}_{t}^{\dagger}
\right\vert \boldsymbol{r}_{t}\sigma\right\rangle ,
\label{eq:local_sf_operator}
\end{equation}
where $\hat{G}^{R}(\omega)=(\omega^{+}-H)^{-1}$ is the full retarded Green's
function, $\omega^{+}=\omega+i\eta$, and $\eta$ is a phenomenological
broadening. The tip-system coupling is 
\begin{equation}
\hat{V}_{t} = \sum_{\sigma} \left[ t_{d}\left\vert \boldsymbol{r}_{t}
\sigma\right\rangle \left\langle d\sigma\right\vert + \sum_{\boldsymbol{k}}
t_{\boldsymbol{k}\sigma} \left\vert \boldsymbol{r}_{t}\sigma\right\rangle
\left\langle \boldsymbol{k}\sigma\right\vert \right] +\mathrm{H.c.},
\label{eq:tip_system_coupling}
\end{equation}
where $\mathrm{H.c.}$ denotes the Hermitian conjugate, and the tunneling
amplitudes are 
\begin{equation}
\left\{ 
\begin{array}{c}
\displaystyle t_{d} = \mathcal{A}_{it} \langle\boldsymbol{r}_{t}\sigma
|d\sigma\rangle, \\[6pt] 
\displaystyle t_{\boldsymbol{k}\sigma} = \mathcal{A}_{at} \langle 
\boldsymbol{r}_{t}\sigma|\boldsymbol{k}\sigma\rangle.%
\end{array}
\right.  \label{eq:method_tip_tunneling_amplitudes}
\end{equation}
Both the tip and impurity orbitals are modeled as isotropic and spin
independent, so the direct tip-impurity tunneling amplitude $t_{d}$ is taken
to be spin independent.

Using Eq.~(\ref{eq:tip_system_coupling}), we separate the local spectral
function into an impurity-free background and an impurity-induced modulation,

\begin{equation}
A_{\sigma}(d,\theta) = A_{0}(d,\theta) - 2\func{Im}\left[ N_{\sigma
}G_{d\sigma,d\sigma}^{R}(E_{F}) \right] .  \label{eq:local_sf_def}
\end{equation}
The background term is 
\begin{equation}
A_{0}(d,\theta) = -2|\mathcal{A}_{at}|^{2} \func{Im} \left[ g_{s\sigma}^{R}(%
\boldsymbol{r}_{t},\boldsymbol{r}_{t};E_{F}) \right] .
\label{eq:background_sf_def}
\end{equation}
Here $g_{s\sigma}^{R}(\boldsymbol{r},\boldsymbol{r}^{\prime};\omega)$ and $%
G_{d\sigma,d\sigma}^{R}(\omega)$ denote the bare substrate Green's function
and the impurity Green's function, respectively (see Appendix~\ref%
{app:eom_derivation} for details). The tip position is measured relative to
the impurity as $\boldsymbol{r}_{t}-\boldsymbol{r}_{d}=d(\cos\theta,\sin%
\theta)$, where $d$ is the tip-impurity separation and $\theta$ is the
corresponding azimuthal angle.

The background term $A_{0}$ is spin independent because it involves only the
coincident-point substrate Green's function $g_{s\sigma }^{R}(\boldsymbol{r}%
_{t},\boldsymbol{r}_{t};E_{F})$. Since there is no finite propagation
distance at coincident points, the direction-dependent altermagnetic
anisotropy drops out. By contrast, the impurity-induced modulation contains
the finite-distance substrate Green's function through 
\begin{equation}
\begin{aligned} N_{\sigma}={}& \left[
t_{d}+\mathcal{A}_{at}\mathcal{A}_{ai}^{\ast}
g_{s\sigma}^{R}(\boldsymbol{r}_{t},\boldsymbol{r}_{d};E_{F}) \right] \\
&\times \left[ t_{d}^{\ast}+\mathcal{A}_{at}^{\ast}\mathcal{A}_{ai}
g_{s\sigma}^{R}(\boldsymbol{r}_{d},\boldsymbol{r}_{t};E_{F}) \right].
\end{aligned}  \label{eq:N_sigma_def}
\end{equation}

In our contact-potential geometry, the tunneling paths do not form a closed
loop enclosing a finite magnetic flux. Therefore, no gauge-invariant
Aharonov-Bohm phase appears. We adopt a phase convention in which the
open-path Peierls phases in the tunneling amplitudes and substrate Green's
functions are omitted. With this convention, the finite-distance substrate
Green's function at the Fermi energy is given by (see Appendices~\ref%
{app:local_spectral_function} and \ref{app:substrate_gf} for details) 
\begin{align}
& g_{s\sigma}^{R}(\boldsymbol{r}_{t},\boldsymbol{r}_{d};E_{F})=g_{s\sigma
}^{R}(\boldsymbol{r}_{d},\boldsymbol{r}_{t};E_{F})=  \notag \\
& \left\{ 
\begin{array}{l}
\displaystyle-\frac{1}{\sqrt{1-\mathcal{J}^{2}}}\frac{im}{2\hbar^{2}}
H_{0}^{(1)}\left( \kappa_{\sigma}k_{F}d\right) , \\[8pt] 
\displaystyle-\frac{1}{\sqrt{1-\mathcal{J}^{2}}}\frac{m}{2\pi\hbar^{2}}
\Gamma_{\mathrm{E}}\left( \frac{1}{2}-\zeta^{+}\right) \frac{W_{\zeta^{+}
,0}\left( R_{\sigma}^{2}\right) }{R_{\sigma}}.%
\end{array}
\right.  \label{eq:substrate_gf_summary}
\end{align}
The first and second lines correspond to the zero-field and Landau-quantized
regimes, respectively. Here $H_{0}^{(1)}(z)$ is the Hankel function of the
first kind, $\Gamma_{\mathrm{E}}(z)$ is the Euler Gamma function, and $%
W_{a,b}(z)$ is the Whittaker function. The spin-dependent anisotropy factor
in Eq.~(\ref{eq:substrate_gf_summary}) is [see Fig.~\ref{fig:setup_model}%
(c)] 
\begin{equation}
\kappa_{\sigma}(\theta)=\sqrt{\frac{1-\sigma\mathcal{J}\sin2\theta }{1-%
\mathcal{J}^{2}}},
\end{equation}
with $k_{F}=\sqrt{2mE_{F}/\hbar^{2}}$ the Fermi wave vector. In the
Landau-quantized regime, we further define the modified cyclotron frequency 
\begin{equation}
\omega_{L}=\sqrt{1-\mathcal{J}^{2}}\frac{eB}{mc},
\end{equation}
and introduce 
\begin{equation}
\left\{ 
\begin{array}{c}
\displaystyle\zeta^{+}=\frac{E_{F}+i\eta}{\hbar\omega_{L}}, \\[10pt] 
\displaystyle R_{\sigma}^{2}=\frac{(\kappa_{\sigma}k_{F}d)^{2}}{4\zeta},%
\end{array}
\right.
\end{equation}
with $\zeta=\zeta^{+}|_{\eta=0}$.

The impurity retarded Green's function is 
\begin{equation}
G_{d\sigma ,d\sigma }^{R}(\omega )=\frac{1}{\omega ^{+}-\mathcal{E}_{\sigma
}-\Sigma _{\sigma }^{R}\left( \omega \right) },  \label{eq:impurity_gf_bare}
\end{equation}%
where the impurity self-energy is 
\begin{align}
\Sigma _{\sigma }^{R}\left( \omega \right) & =\sum_{\boldsymbol{k}}V_{%
\boldsymbol{k}\sigma }g_{\boldsymbol{k}\sigma ,\boldsymbol{k}\sigma
}^{R}\left( \omega \right) V_{\boldsymbol{k}\sigma }^{\ast }  \notag \\
& =\left\vert \mathcal{A}_{ai}\right\vert ^{2}g_{s\sigma }^{R}(\boldsymbol{r}%
_{d},\boldsymbol{r}_{d};\omega ).  \label{eq:impurity_self_energy}
\end{align}

The coincident-point substrate Green's function in Eq.~(\ref%
{eq:impurity_self_energy}) contains an ultraviolet divergence. After
regularizing it, Eq.~(\ref{eq:impurity_gf_bare}) at the Fermi energy becomes
(see Appendix~\ref{app:impurity_renormalization} for details)

\begin{equation}
G_{d\sigma,d\sigma}^{R}(E_{F})=\frac{1}{\Delta_{\sigma}-\Lambda+i\Gamma}.
\label{eq:impurity_gf_renormalized}
\end{equation}
Here $\Delta_{\sigma}=E_{F}-\left( \xi+\sigma\varepsilon_{d}\right) $ is the
spin-dependent impurity-level detuning, with $\xi$ the renormalized impurity
level. The quantities $\Lambda$ and $\Gamma$ denote the regularized
impurity-level shift and linewidth, respectively. They are given by 
\begin{equation}
\Lambda=\left\{ 
\begin{array}{l}
\displaystyle0, \\[10pt] 
\displaystyle\frac{1}{\sqrt{1-\mathcal{J}^{2}}}\frac{m|\mathcal{A}_{ai}|^{2} 
}{2\pi\hbar^{2}}\left\{ \func{Re}\left[ \psi\left( 1/2-\zeta ^{+}\right) %
\right] -\ln\zeta\right\} ,%
\end{array}
\right.  \label{eq:impurity_level_shift}
\end{equation}
and 
\begin{equation}
\Gamma=\left\{ 
\begin{array}{l}
\displaystyle\frac{1}{\sqrt{1-\mathcal{J}^{2}}}\frac{m|\mathcal{A}_{ai}|^{2} 
}{2\hbar^{2}}, \\[10pt] 
\displaystyle-\frac{1}{\sqrt{1-\mathcal{J}^{2}}}\frac{m|\mathcal{A}%
_{ai}|^{2} }{2\pi\hbar^{2}}\func{Im}\left[ \psi\left( 1/2-\zeta^{+}\right) %
\right] ,%
\end{array}
\right.  \label{eq:impurity_level_broadening}
\end{equation}
with $\psi(z)$ the digamma function.

For real $\mathcal{A}_{at}$, $\mathcal{A}_{ai}$, and $t_{d}$, Eq.~(\ref%
{eq:local_sf_def}) takes the generalized Fano form 
\begin{align}
& A_{\sigma }(d,\theta )=A_{0}(d,\theta )\times  \notag \\
& \left\{ 1+\left[ \func{Re}Z_{\sigma }\right] ^{2}\left[ 1-\frac{\left[
\left( \Delta _{\sigma }-\Lambda \right) /\Gamma +q_{\sigma }\right] ^{2}}{%
\left( \Delta _{\sigma }-\Lambda \right) ^{2}/\Gamma ^{2}+1}\right] \right\}
,  \label{eq:local_fano_general}
\end{align}%
where the (spin-dependent, anisotropic) Fano $q$ factor is 
\begin{equation}
q_{\sigma }=\tan \left[ \arg Z_{\sigma }\right] .
\end{equation}%
Here, 
\begin{equation}
Z_{\sigma }=\bar{g}_{s\sigma }^{R}(d,\theta )+\lambda ,
\end{equation}%
where 
\begin{align}
& \bar{g}_{s\sigma }^{R}(d,\theta )=-\frac{g_{s\sigma }^{R}(\boldsymbol{r}%
_{t},\boldsymbol{r}_{d};E_{F})}{\func{Im}\left[ g_{s\sigma }^{R}(\boldsymbol{%
r}_{t},\boldsymbol{r}_{t};E_{F})\right] }=  \notag \\
& \left\{ 
\begin{array}{l}
\displaystyle-iH_{0}^{(1)}\left( \kappa _{\sigma }k_{F}d\right) , \\[8pt] 
\displaystyle\dfrac{\Gamma _{\mathrm{E}}\left( 1/2-\zeta ^{+}\right) }{\func{%
Im}\left[ \psi \left( 1/2-\zeta ^{+}\right) \right] }\dfrac{W_{\zeta
^{+},0}\left( R_{\sigma }^{2}\right) }{R_{\sigma }},%
\end{array}%
\right.  \label{eq:normalized_substrate_gf_summary}
\end{align}%
is the dimensionless substrate Green's function. The first and second lines
correspond to the zero-field and Landau-quantized regimes, respectively. The
dimensionless parameter 
\begin{equation}
\lambda =-\frac{t_{d}}{\mathcal{A}_{at}\mathcal{A}_{ai}^{\ast }\func{Im}%
\left[ g_{s\sigma }^{R}(\boldsymbol{r}_{t},\boldsymbol{r}_{t};E_{F})\right] }
\end{equation}%
parametrizes the direct tip-impurity channel relative to the
substrate-mediated channel. Like $A_{0}$, the quantities $\lambda $, $%
\Lambda $, and $\Gamma $ are spin independent because they involve only
coincident-point substrate Green's functions, for which no finite
propagation distance enters. The spin dependence of the Fano $q$ factor
therefore originates from the finite-distance Green's function $\bar{g}%
_{s\sigma }^{R}(d,\theta )$.

\section{Results and Discussion}

\begin{figure}[ptb]
\centering
\includegraphics[width=1\linewidth]{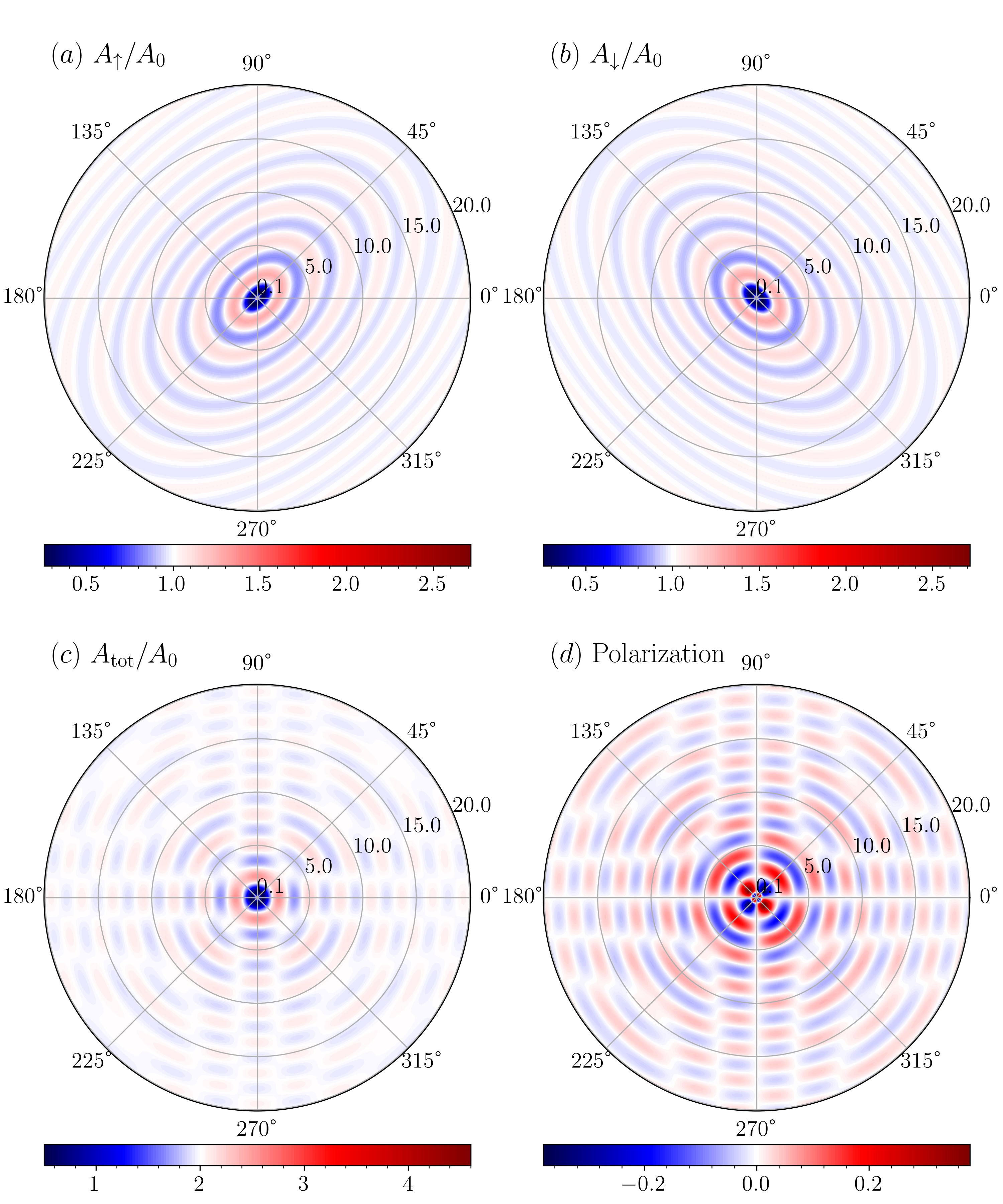}
\caption{(a)--(c) Spin-up, spin-down, and total local spectral functions in
the zero-field regime. The radial and angular coordinates denote $k_{F}d$
and the tip angle $\protect\theta$, respectively. (d) Corresponding spin
polarization. The impurity-level detuning is set to $\Delta/\Gamma=0$.}
\label{fig:zero-field_spatial_sf}
\end{figure}

\begin{figure*}[ptb]
\centering
\includegraphics[width=.9\linewidth]{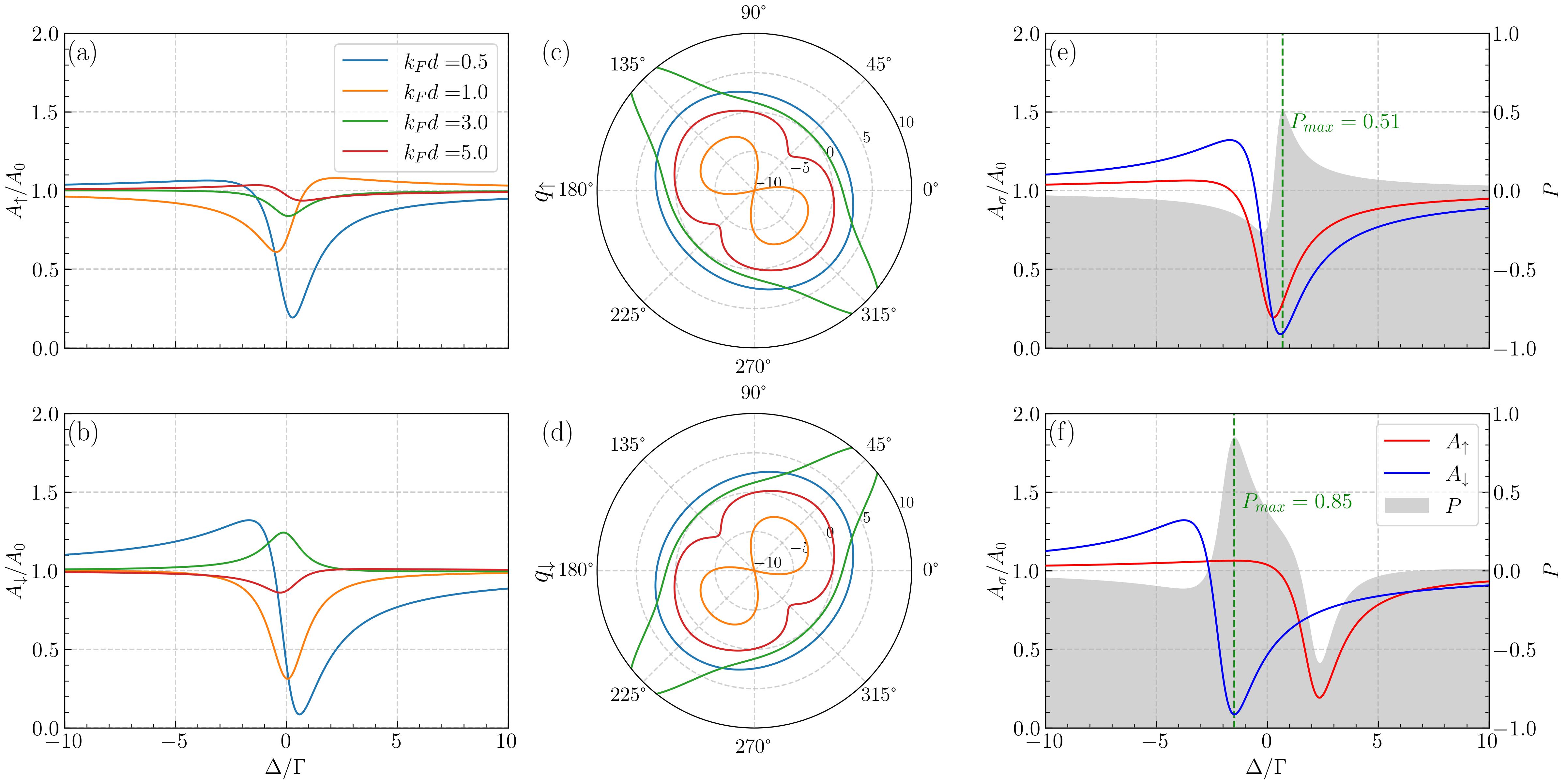}
\caption{(a), (b) Spin-up and spin-down local spectral functions as
functions of the tip-impurity separation $k_{F}d$. (c), (d) Corresponding
Fano $q$ factors. (e) Local spectral functions and spin polarization at $%
k_{F}d=0.5$. (f) Results with an optimized impurity Zeeman splitting $%
\protect\varepsilon_{d}/\Gamma=-2.07$ that maximizes the spin polarization.
For the local-spectral-function panels, the orientation is fixed at $\protect%
\theta=3\protect\pi/4$.}
\label{fig:zero-field_detuning_sf}
\end{figure*}

We begin by specifying the parameters and nondimensionalization scheme. We
focus on a shallow effective altermagnetic pocket, with the Fermi energy $%
E_{F}$ measured from the pocket bottom. Representative estimates give $%
E_{F}\simeq 5.3~\mathrm{meV}$ and $k_{F}^{-1}\simeq 3.8~\mathrm{nm}$. These
values place the impurity-induced spectral patterns on an STM-accessible
real-space length scale and allow moderate Landau-level filling factors at
experimentally accessible magnetic fields. Based on the estimates summarized
in Appendix~\ref{app:parameter_estimates}, we set the altermagnetic
parameter to $\mathcal{J}=0.4$, which lies within the stable regime $%
\mathcal{J}<1$ and produces a sizable spin splitting.

We normalize the local spectral function $A_{\sigma}$ by its impurity-free
counterpart $A_{0}$. Substrate energy scales, including the phenomenological
broadening $\eta$ and the cyclotron energy $\hbar\omega_{L}$, are measured
in units of $E_{F}$, while lengths are expressed in units of $k_{F}^{-1}$.
Impurity-related energy scales, such as the detuning $\Delta_{\sigma}$ and
the level shift $\Lambda$, are measured in units of the impurity linewidth $%
\Gamma$. Varying the dimensionless detuning $(\Delta_{\sigma}-\Lambda)/%
\Gamma $ is therefore used to represent sweeping the energy through the
impurity resonance. The impurity Zeeman splitting is set to zero unless
stated otherwise. Thus, we use the spin-independent detuning $%
\Delta=E_{F}-\xi$ rather than $\Delta_{\sigma}$ in most cases.

We first examine the limit of vanishing direct tip-impurity tunneling, $%
\lambda\to 0$. This limit is realized when the tip-impurity separation is
sufficiently large that the direct overlap between the STM tip and the
impurity orbital becomes negligible. In this regime, the STM response is
dominated by the substrate-mediated channel, modified by resonant scattering
from the impurity:

\begin{equation}
A_{\sigma }(d,\theta )\simeq -2\left\vert \mathcal{A}_{at}\right\vert ^{2}%
\func{Im}\left[ G_{s\sigma }^{R}(\boldsymbol{r}_{t},\boldsymbol{r}_{t};E_{F})%
\right] ,
\end{equation}%
where $G_{s\sigma }^{R}$ is the full retarded Green's function projected
onto the substrate subspace (see Appendix~\ref{app:local_spectral_function}
for details), 
\begin{align}
& G_{s\sigma }^{R}(\boldsymbol{r}_{t},\boldsymbol{r}_{t};\omega )=g_{s\sigma
}^{R}(\boldsymbol{r}_{t},\boldsymbol{r}_{t};\omega )  \notag \\
& \quad +|\mathcal{A}_{ai}|^{2}g_{s\sigma }^{R}(\boldsymbol{r}_{t},%
\boldsymbol{r}_{d};\omega )G_{d\sigma ,d\sigma }^{R}(\omega )g_{s\sigma
}^{R}(\boldsymbol{r}_{d},\boldsymbol{r}_{t};\omega ).
\end{align}%
The first and second terms describe the bare substrate contribution and the
impurity-induced correction to the substrate Green's function, respectively.

\begin{figure*}[ptb]
\centering
\includegraphics[width=.9\linewidth]{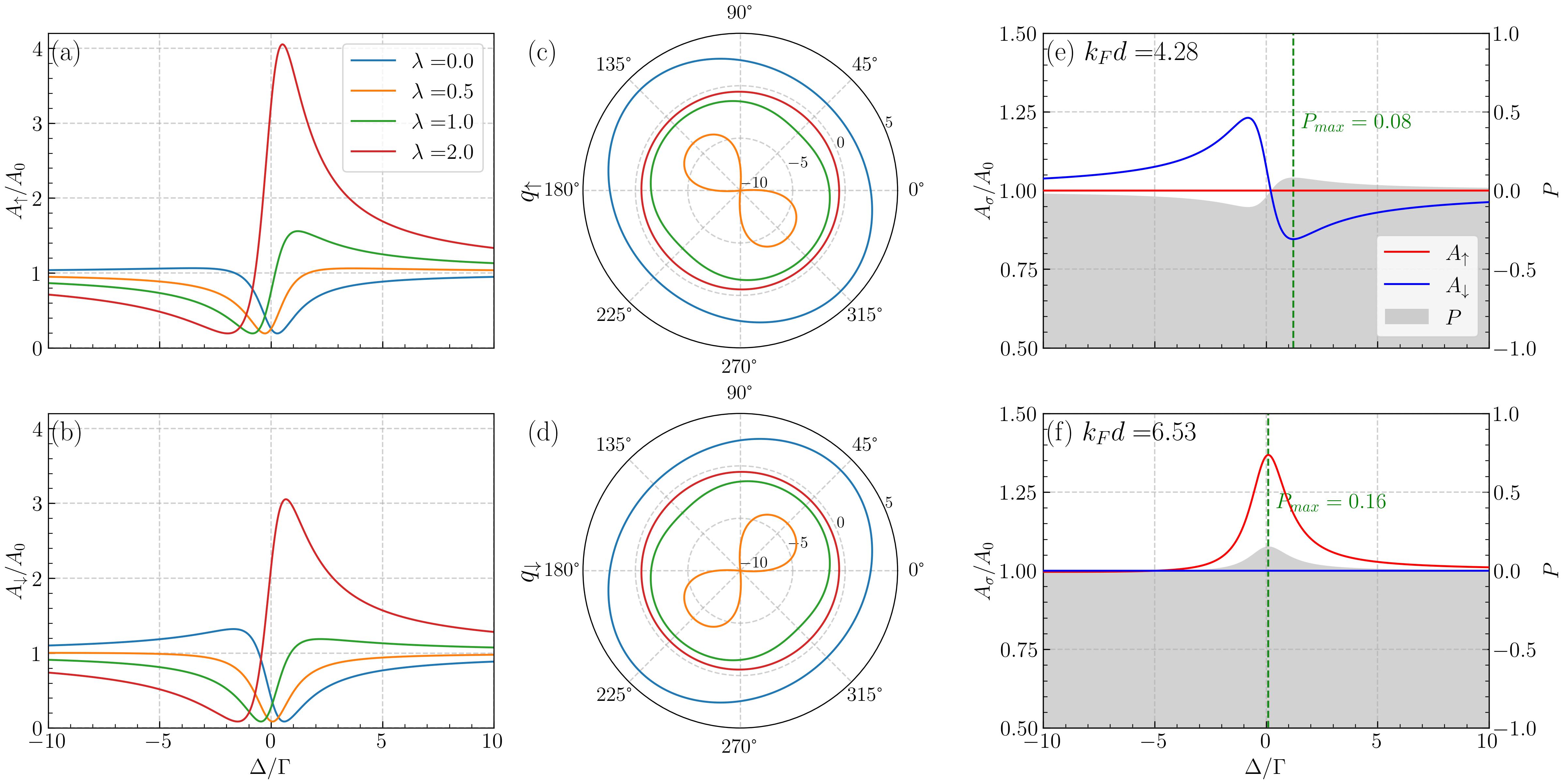}
\caption{(a), (b) Spin-up and spin-down local spectral functions as
functions of the direct tunneling parameter $\protect\lambda$. (c), (d)
Corresponding Fano $q$ factors. (e) Local spectral functions and spin
polarization at $k_{F}d=4.28$ and $\protect\lambda=0.34$, where the spin-up
branch satisfies the interference-zero condition. (f) Corresponding results
for the spin-down interference zero, with $k_{F}d=6.53$ and $\protect\lambda %
=0.34$. For panels (a)--(d), $k_{F}d=0.5$ and $\protect\theta=3\protect\pi/4$
.}
\label{fig:zero-field_direct_tip_impurity}
\end{figure*}

In the zero-field regime, $A_{\sigma }$ develops a spin-resolved beating
pattern, as shown in Fig.~\ref{fig:zero-field_spatial_sf}. This behavior is
consistent with earlier results for $\delta $-type scattering potential \cite%
{Chen2024ImpurityScatteringAltermagnets,Gondolf2025LocalSignaturesAltermagnetism,Hu2025QuasiparticleInterference,Sukhachov2024ImpurityInducedFriedel}%
. The spatial structure originates from the dimensionless substrate Green's
function $\bar{g}_{s\sigma }^{R}(\boldsymbol{r}_{t},\boldsymbol{r}_{d})$
[see Eq.~(\ref{eq:normalized_substrate_gf_summary})]. As in the
non-altermagnetic case, its radial dependence is governed by the Hankel
function $H_{0}^{(1)}(\kappa_{\sigma }k_{F}d)$, so $A_{\sigma }$ exhibits
oscillatory spatial decay for each direction $\theta $. At large distances,
the Hankel dependence leads to a Friedel-like oscillation, as the
impurity-induced correction $\delta A_{\sigma }\equiv A_{\sigma }-A_{0}$
obeys the asymptotic form 
\begin{equation}
\lim_{d\rightarrow \infty }\delta A_{\sigma }(d,\theta )=-\frac{1}{\sqrt{1- 
\mathcal{J}^{2}}}\frac{2m}{\pi \hbar ^{2}\mathcal{R}}\frac{\sin \!\left(
2\kappa_{\sigma }k_{F}d+\delta \right) }{\kappa_{\sigma }k_{F}d},
\end{equation}%
with $\sin \delta =\Delta /(\mathcal{R}\Gamma )$, $\cos \delta =1/\mathcal{R}
$, and $\mathcal{R}=\sqrt{1+\Delta ^{2}/\Gamma ^{2}}$. This expression
describes a sinusoidal oscillation with an algebraically decaying $1/d$
envelope. Both the oscillation period, 
\begin{equation}
L_{\sigma }(\theta )=\frac{\pi }{\kappa_{\sigma }(\theta )k_{F}},
\end{equation}
and the angular dependence of the amplitude are controlled by the
altermagnetic anisotropy factor $\kappa_{\sigma }(\theta)$. Hence, the
spatial period provides a route to estimate the strength of the
altermagnetic splitting: because $\mathcal{J}$ is encoded in $\kappa_{\sigma
}$, it can be extracted by comparing the two spin-resolved periods, namely 
\begin{equation}
\mathcal{J}=\frac{(L_{\uparrow }/L_{\downarrow})^{2}-1}{\sin 2\theta
\,[(L_{\uparrow }/L_{\downarrow})^{2}+1]}.  \label{eq:J_extraction}
\end{equation}

The same anisotropy also controls the full real-space pattern beyond the
asymptotic period. Specifically, the spatial pattern of $\delta A_{\sigma }$
is distorted from concentric circles into concentric ellipses whose
principal axes lie along the diagonal ($\theta =\pi /4,3\pi /4$) directions.
Because $\delta A_{\uparrow }$ and $\delta A_{\downarrow }$ are related by $%
\hat{C}_{4}\hat{T}$ symmetry, the major axis of the $\delta A_{\uparrow }$
ellipses coincides with the minor axis of the $\delta A_{\downarrow }$
ellipses, and vice versa. This mismatch in oscillation periods generates a
beating pattern in the total correction $\delta A_{\mathrm{tot}}\equiv
\delta A_{\uparrow }+\delta A_{\downarrow }$ along the diagonal directions.
Away from these directions, beating also arises from the difference between $%
\kappa_{\uparrow }$ and $\kappa_{\downarrow }$. Along the horizontal ($%
\theta =0$) and vertical ($\theta =\pi /2$) directions, the oscillation
periods of the two spin channels coincide, as required by $\hat{C}_{4}\hat{T}
$ symmetry.

The agreement with previous potential-scattering results is rooted in the
underlying scattering $T$-matrix structure. From this perspective, the
resonant impurity can be mapped onto a regularized $\delta$-type scattering
potential \cite{rusin2018theory}, whose effective strength scales as $%
V_{r}\sim|\mathcal{A}_{ai}|^{2}/\Delta^{+}$ and is therefore controlled by
the impurity-level detuning (see Appendix~\ref{app:scatterer_mapping}).
Retaining the explicit impurity degree of freedom, rather than replacing it
with a scattering potential, makes the resonant-scattering mechanism
explicit. It exposes the Fano-type line-shape asymmetry [see Eq.~(\ref%
{eq:local_fano_general})] and identifies the impurity-level detuning as a
control parameter for the real-space response.

In the $\lambda\to 0$ limit, the Fano $q$ factor, which controls the
asymmetry of $A_\sigma$, reduces to 
\begin{equation}
q_{\sigma} = \tan\!\left\{ \arg\!\left[\bar{g}_{s\sigma}^{R}(d,\theta)\right]
\right\}.
\end{equation}
This result shows that the line shape can remain asymmetric even without a
direct tip-impurity tunneling channel. In this limit, the asymmetry is a
phase-sensitive effect encoded in the complex substrate Green's function 
\cite{Ujsaghy2000FanoSTM}.

This conclusion is also evident from the Lippmann--Schwinger equation. In
that picture, the substrate-projected scattering state consists of an
incoming plane-wave component and a scattered component. The resonant
behavior enters the latter through the impurity Green's function. When the
local spectral function is evaluated at the tip position, the cross term
between the incoming and resonantly scattered components produces the
asymmetric Fano profile (see Appendix~\ref{app:lse_asymmetry}).

This phase sensitivity also links the spin-resolved line shapes to the STM
tip position. Since $q_{\sigma}$ is determined by the phase of the substrate
Green's function between the impurity and the tip, moving the tip alters the
propagation phase and therefore modifies the Fano asymmetry. Because $%
\kappa_{\uparrow}\neq\kappa_{\downarrow}$ at generic angles, the two spin
channels accumulate different phases and thus acquire distinct Fano factors, 
$q_{\uparrow}\neq q_{\downarrow}$. This leads to a mismatch between the
spin-resolved line shapes in the energy domain, as shown in Fig.~\ref%
{fig:zero-field_detuning_sf}(a)--(d). The same spatial dependence provides a
complementary route to extract $\mathcal{J}$. At large distances, one has 
\begin{equation}
\lim_{d\rightarrow \infty }q_{\sigma }(d,\theta) = -\cot\left(
\kappa_{\sigma }k_{F}d-\frac{\pi }{4}\right) .
\end{equation}
Therefore, $q_{\sigma}$ has the same spatial period $L_{\sigma }(\theta)$ as
the impurity-induced oscillation. By fitting energy-domain STS line shapes
measured at different tip positions, one obtains the position-dependent Fano
factor $q_{\sigma}(d,\theta)$, whose spin-resolved periods provide an
alternative extraction of $\mathcal{J}$ through Eq.~(\ref{eq:J_extraction}).
Compared with fitting a single fixed-energy real-space pattern, this global
spectroscopic extraction uses more fitting constraints and can therefore
reduce the error in $L_{\sigma}(\theta)$ and the uncertainty in $\mathcal{J}$%
.

The line-shape mismatch also enhances the spin polarization of the local
spectral function, 
\begin{equation}
P\equiv \frac{A_{\uparrow}-A_{\downarrow}} {A_{\uparrow}+A_{\downarrow}}.
\end{equation}
Intuitively, the enhancement is most effective when the tip position is
chosen so that the resonant peak of one spin branch aligns closely with the
antiresonant dip of the other. Under such peak-dip alignment, tuning the
impurity-level detuning $\Delta$ into the corresponding energy window
suppresses one spin component while preserving the other, yielding a large
spin contrast and an enhanced $P$.

However, the tip position affects not only the line-shape asymmetry but also
the magnitude of the impurity-induced modulation. This magnitude is
controlled by the prefactor $[\func{Re} Z_\sigma]^2$ in Eq.~(\ref%
{eq:local_fano_general}), which reduces to $\{\func{Re}[\bar{g}%
_{s\sigma}^R(d,\theta)]\}^2$ in the $\lambda\to 0$ limit. This creates a
trade-off: a large prefactor favors a small tip-impurity separation $d$,
whereas a pronounced mismatch between $A_\uparrow$ and $A_\downarrow$, which
is essential for a large $P$, typically develops more efficiently at larger $%
d$. Optimizing the spin polarization by tuning the tip position alone is
therefore difficult. This trade-off can be mitigated by introducing an
additional control parameter, such as an impurity Zeeman splitting $%
\varepsilon_d$. In this way, the tip position can be chosen primarily to
maximize the prefactor, while $\varepsilon_d$ independently tunes the
peak-dip alignment between the two spin channels [see Fig.~\ref%
{fig:zero-field_detuning_sf}(e), (f)].

We next include the direct tip-impurity coupling, which becomes important at
small tip-impurity separations $d$. This coupling introduces a
spin-independent tunneling channel and thereby reduces the contrast between $%
A_\uparrow$ and $A_\downarrow$ [see Fig.~\ref%
{fig:zero-field_direct_tip_impurity}(a)--(d)]. At the same time, it connects
the line-shape asymmetry to the interference between tunneling paths: the
direct tip-impurity channel and the substrate-mediated channel. In the
wide-band approximation, where the real part of $g_{s\sigma}^R(\boldsymbol{r}%
_t,\boldsymbol{r}_d;E_{F})$ is neglected, the $q$ factor simplifies to

\begin{equation}
q_{\sigma}\simeq\frac{\mathcal{A}_{at}\mathcal{A}_{ai}^{\ast}}{t_{d}} \func{%
Im}\left[ g_{s\sigma}^{R}(\boldsymbol{r}_{t},\boldsymbol{r} _{d};E_{F})%
\right] .
\end{equation}
In this approximation, $A_\sigma$ becomes a symmetric Lorentzian in the $%
\lambda\to0$ limit, while at finite $\lambda$ the asymmetry reflects the
relative strength of the two tunneling channels: the direct channel
(strength $t_d$) and the substrate-mediated channel (strength $\mathcal{A}%
_{at}\mathcal{A}_{ai}^{\ast}\,g_{s\sigma}^R$) (see Appendix~\ref%
{app:lse_asymmetry}). Beyond this approximation, however, the direct
tip-impurity channel should be viewed as an additional contributor that
modifies the line-shape asymmetry, rather than as its sole origin. Even in
the $\lambda\to0$ limit, the line shape retains a phase-sensitive asymmetry
inherited from the substrate Green's function.

The direct tip-impurity channel also enables an interference zero in the
impurity-induced correction $\delta A_\sigma$. This zero occurs when $%
Z_{\sigma}=0$, as Eq.~(\ref{eq:local_fano_general}) can be written as 
\begin{equation}
\frac{\delta A_{\sigma}}{A_{0}}=-\func{Im}\left[ \frac{Z_{\sigma} ^{2}}{%
\left( \Delta_{\sigma}-\Lambda\right) /\Gamma+i}\right] .
\end{equation}

Using Eq.~(\ref{eq:normalized_substrate_gf_summary}), the condition $%
Z_\sigma=0$ in the zero-field regime becomes

\begin{equation}
\left\{ 
\begin{array}{c}
J_{0}(\kappa _{\sigma }k_{F}d)=0, \\ 
\lambda =-Y_{0}(\kappa _{\sigma }k_{F}d),%
\end{array}%
\right.
\end{equation}%
where $J_{n}(x)$ and $Y_{n}(x)$ are Bessel functions of the first and second
kinds, respectively. These equations can be solved using the zeros of $%
J_{0}(x)$. For instance, near the second positive zero of $J_{0}\left(
\kappa _{\sigma }k_{F}d\right) $, namely $\kappa _{\sigma }k_{F}d\simeq 5.52$%
, one finds $\lambda \simeq 0.34$. Because $\kappa _{\uparrow }\neq \kappa
_{\downarrow }$ at generic angles, the conditions for $Z_{\uparrow }=0$ and $%
Z_{\downarrow }=0$ are generally mismatched. Thus, by tuning the tip
position and the Fermi level, one can approach the interference zero for one
spin branch while keeping the impurity-induced response finite in the other
[see Fig.~\ref{fig:zero-field_direct_tip_impurity}(e), (f)]. This
spin-selective suppression of $\delta A_{\sigma }$ provides a local
spectroscopic signature of spin-dependent substrate propagation and a
mechanism for enhancing the spin contrast.

\begin{figure}[ptb]
\centering
\includegraphics[width=1\linewidth]{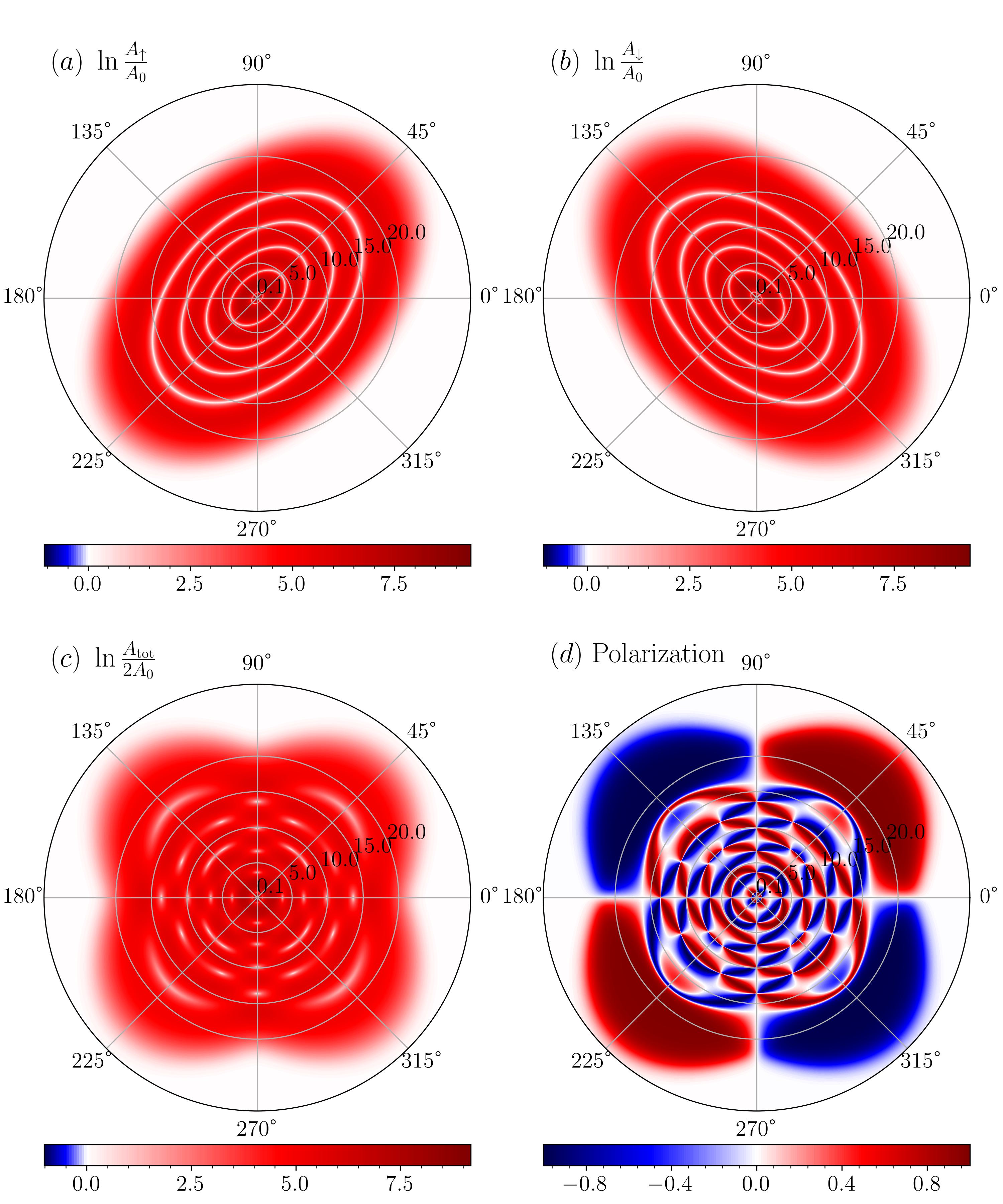}
\caption{(a)--(c) Logarithm of the spin-up, spin-down, and averaged total
local spectral functions, respectively, in the Landau-quantized regime. The
logarithm is plotted because the local spectral function varies over a wide
range. Radial and angular coordinates denote $k_{F}d$ and $\protect\theta$,
respectively. (d) Corresponding spin polarization with $P_{\max}\approx0.99$%
. Parameters are $\Delta/\Gamma=0$, $\protect\eta=10^{-3}E_{F}$, and $%
\protect\zeta=5$, corresponding to the Fermi level lying midway between the
fifth and sixth Landau levels.}
\label{fig:landau_spatial_sf}
\end{figure}

\begin{figure}[ptb]
\centering
\includegraphics[width=1\linewidth]{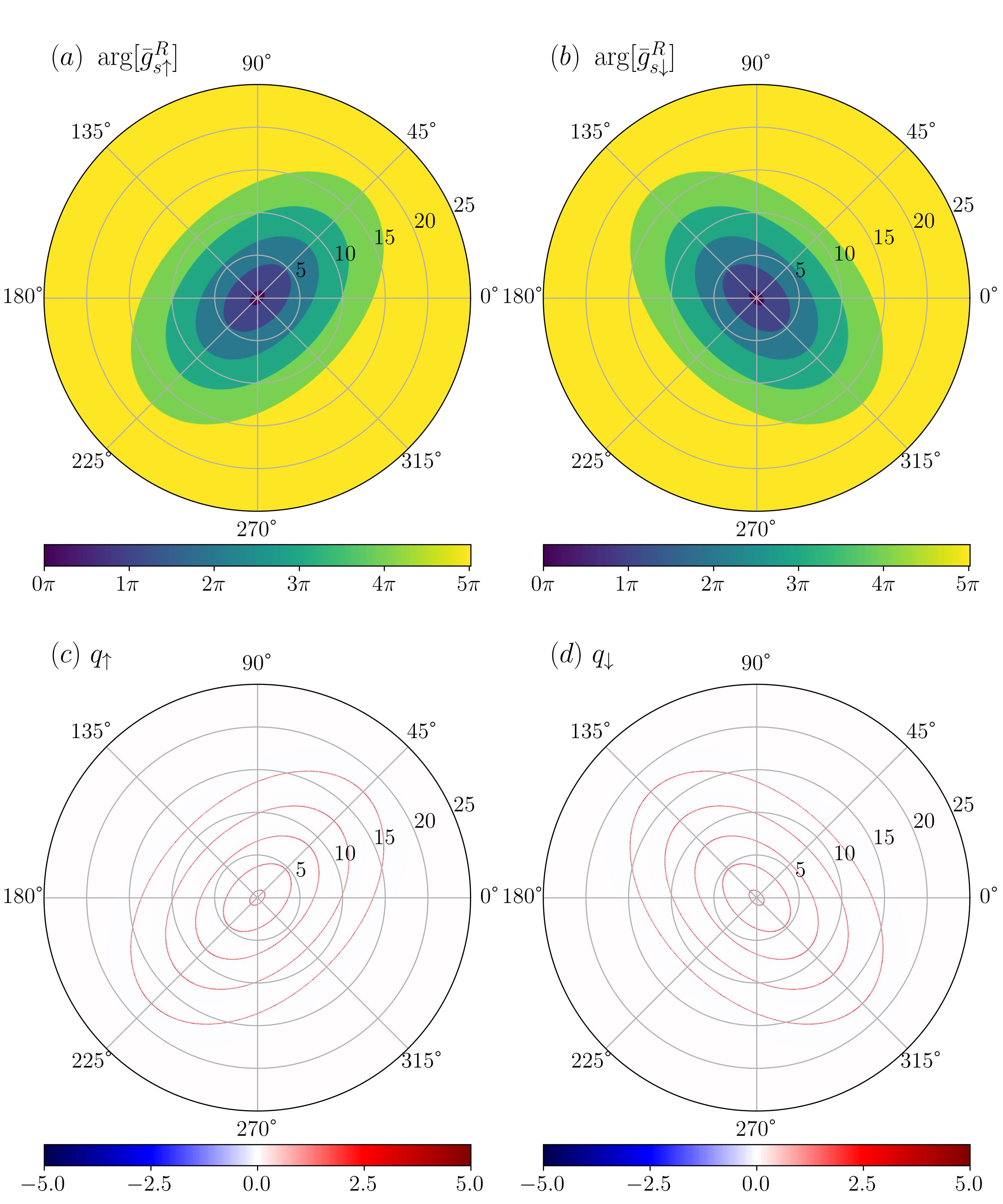}
\caption{Phase of the dimensionless substrate Green's function $\bar{g}_{s 
\protect\sigma}^{R}(\boldsymbol{r}_{t},\boldsymbol{r}_{d})$ for the (a)
spin-up and (b) spin-down channels. (c), (d) Corresponding Fano $q$ factors.
The radial and angular coordinates denote $k_{F}d$ and $\protect\theta$,
respectively.}
\label{fig:landau_fano_phase}
\end{figure}

We now examine the Landau-quantized regime. In the $\lambda\to0$ limit, $%
A_\sigma$ does not exhibit simple oscillations. Instead, it develops a nodal
pattern in real space [see Fig.~\ref{fig:landau_spatial_sf}(a) and (b)].
This structure originates from the spatial dependence of the dimensionless
substrate Green's function $\bar{g}_{s\sigma}^R(\boldsymbol{r}_t,\boldsymbol{%
r}_d)$, which is governed by the Whittaker function $W_{\zeta^+,0}(R_%
\sigma^2)$. For a real $\zeta$ in the interval $(n-1/2,n+1/2)$, that is,
when the Fermi level lies between the $n$th and $(n+1)$th Landau levels, the
function $W_{\zeta,0}(R_\sigma^2)$ exhibits $n$ nodes along the positive $d$
axis. A finite broadening $\eta$, which replaces $\zeta$ by $\zeta^+$,
preserves this nodal pattern qualitatively but turns the exact zeros into
shallow minima. At large tip-impurity separations, the Whittaker function
decays rapidly because of Landau confinement, so $\delta A_\sigma$ is
suppressed and $A_\sigma$ approaches the impurity-free background.

Since $R_\sigma$ contains the anisotropy factor $\kappa_\sigma(\theta)$, the
nodal pattern of $A_\sigma$ forms a set of concentric ellipses whose
principal axes are interchanged between the two spin channels. This
spin-dependent deformation, combined with the nodal structure, produces a
large spin polarization $P$ [see Fig.~\ref{fig:landau_spatial_sf}(d)]. The
relevant mechanism is spatial overlap mismatch: a nodal region of one spin
branch can coincide with a non-nodal region of the other. In such regions,
the impurity-induced modulation is suppressed in one spin channel while
remaining finite in the opposite channel, yielding a large local spin
polarization. Tip positions near these spin-selective nodal regions
therefore give a large spin polarization in the local tunneling signal.

The real-space spin contrast also depends on the energy dependence of the
impurity resonance. Although the local spectral function still follows the
Fano form of Eq.~(\ref{eq:local_fano_general}), the Landau-quantized regime
differs qualitatively from the zero-field regime in its phase structure. In
most spatial regions, $\delta A_\sigma$ exhibits a nearly symmetric
Lorentzian line shape rather than a strongly asymmetric Fano profile. This
behavior originates from the spatial dependence of $\arg[\bar{g}_{s\sigma}^R(%
\boldsymbol{r}_t,\boldsymbol{r}_d)]$, which is again governed by the
Whittaker function $W_{\zeta^+,0}(R_\sigma^2)$.

Figure~\ref{fig:landau_fano_phase} shows that the phase of $\bar{g}%
_{s\sigma}^R(\boldsymbol{r}_t,\boldsymbol{r}_d)$ exhibits a step-like
dependence on the tip-impurity separation. Starting from $\lim_{\boldsymbol{r%
}_t\to\boldsymbol{r}_d} \arg[\bar{g}_{s\sigma}^R(\boldsymbol{r}_t,%
\boldsymbol{r}_d)] = 0$, the phase remains nearly constant as $d$ increases,
except near the amplitude minima, where it jumps abruptly by $\pi$. This
behavior is characteristic of the Whittaker function: each node in the
amplitude is accompanied by a $\pi$ phase jump. Consequently, the Fano
factor $q_\sigma$ stays close to zero in most spatial regions, and the local
spectral function remains nearly symmetric in energy. A finite direct
tip-impurity coupling ($\lambda \neq 0$) mainly shifts the real part of $%
Z_\sigma$ and therefore does not qualitatively alter this conclusion, except
near amplitude-minima regions.

The near-Lorentzian character implies that the magnitude of the
spin-selective response is controlled not only by the spatial nodal
mismatch, but also by the alignment between the impurity resonance and the
Fermi energy. In the Landau-quantized regime, the energy shift $\Lambda$ of
the impurity level [Eq.~(\ref{eq:impurity_level_shift})] can detune the
resonance away from $E_{F}$, suppressing $\delta A_\sigma$. Conversely,
tuning the impurity level, for example by gate voltage or doping, can
restore the resonance condition and enhance the impurity-induced modulation,
thereby amplifying the contrast between nodal and non-nodal regions.

\begin{figure}[ptb]
\centering
\includegraphics[width=1\linewidth]{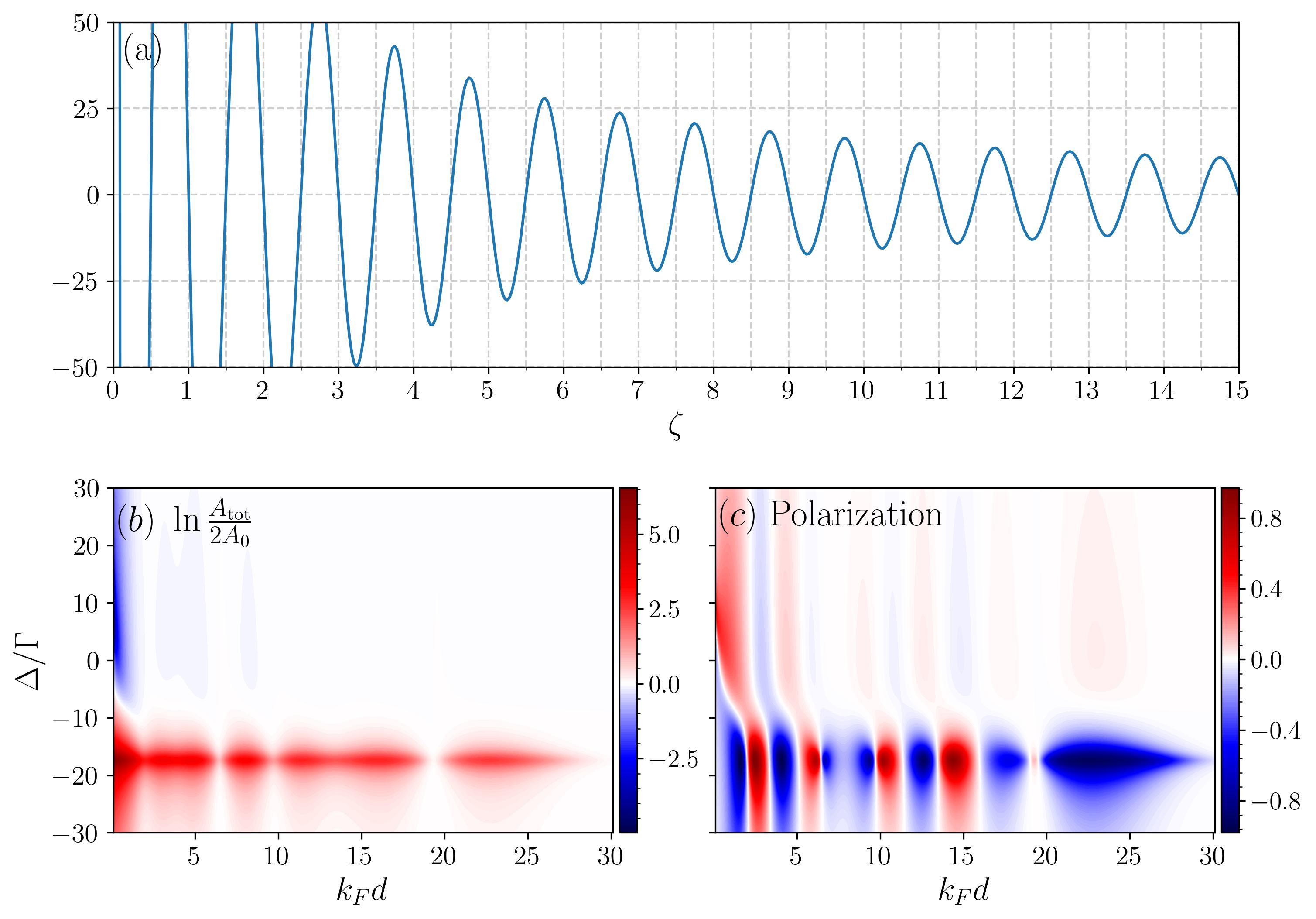}
\caption{(a) Dimensionless impurity-level shift $\Lambda/\Gamma$ as a
function of the Landau-level filling factor $\protect\zeta$. (b) Logarithm
of the averaged total local spectral function and (c) the corresponding spin
polarization as functions of $k_{F}d$ and $\Delta/\Gamma$. In panels (b) and
(c), $\protect\theta=3\protect\pi/4$ is set, and $\protect\zeta=5.4$ is
chosen as a representative case where the impurity-level energy shift is
appreciable.}
\label{fig:landau_gate_tuning}
\end{figure}

Figure~\ref{fig:landau_gate_tuning} illustrates this resonance-alignment
effect. When the impurity resonance is tuned close to the Fermi energy, $%
\delta A_\sigma$ is enhanced and can become the dominant contribution to $%
A_\sigma$. In this regime, the spatial contrast between nodal and non-nodal
regions is amplified, leading to a larger spin-resolved local spectral
function difference $|A_\sigma - A_{\bar{\sigma}}|$ and consequently an
enhanced spin polarization. The impurity detuning $\Delta$ and the
Landau-level filling factor $\zeta$ therefore provide control parameters for
optimizing the spin-selective response.

\section{Conclusion}

We have studied the spin-resolved local spectral function of a resonant
impurity coupled to a two-dimensional $d$-wave altermagnetic substrate.
Using a Green's-function formalism, we have shown that the spin-dependent
altermagnetic anisotropy is encoded in the finite-distance substrate Green's
function connecting the impurity and the tip. In the zero-field regime, the
resonant impurity gives rise to a dual Fano resonance. In the absence of
direct tip-impurity coupling, the impurity resonance is encoded in the
perturbed substrate continuum, and the Fano asymmetry is controlled by the
propagation phase of the substrate Green's function. This Fano profile
produces spin-mismatched line shapes and allows the altermagnetic splitting
parameter $\mathcal{J}$ to be estimated. In the presence of direct
tip-impurity coupling, the response crosses over to a path-interference
(between the direct tip-impurity tunneling pathway and the
substrate-mediated tunneling pathway) mechanism, enabling spin-selective
tunneling by tuning the Fermi energy and the tip position. In the
Landau-quantized regime, the dominant signature is instead the real-space
nodal structure: the local spectral function develops spin-dependent nodal
patterns, and the mismatch between nodal and non-nodal regions of opposite
spin channels gives rise to a large local spin contrast (up to $P\approx 0.99$ for $\hbar\omega_L=5\times10^3\eta$). This spin contrast
can be strongly enhanced when the impurity resonance is aligned with the
Fermi energy. These results establish resonant-impurity STM/STS as a
phase-sensitive local probe of altermagnetic band anisotropy.

\begin{acknowledgments}
We would like to thank Z.-W.-Y. Chen and J.-J. Lin for helpful discussions.
This work was supported by the National Natural Science Foundation of China
(12574032, 12174032), the National Natural Science Foundation of
China-Research Grants Council (11861161002), and the National Key Research
and Development Program of China (2017YFA0303400).
\end{acknowledgments}

\appendix

\setcounter{figure}{0} \renewcommand{\thefigure}{S\arabic{figure}}

\section{Equation-of-motion derivation of the Green's functions}

\label{app:eom_derivation}

We collect here the equation-of-motion (EOM) identities used in the main
text and in Appendix~\ref{app:local_spectral_function}. Throughout this
appendix, we set $\hbar=1$, so that energies and frequencies are used
interchangeably. The retarded Green's function is defined as 
\begin{equation}
G_{a,b}^{R}\left( t,t^{\prime}\right) =-i\theta\left( t-t^{\prime}\right)
\left\langle \left\{ c_{a}\left( t\right) ,c_{b}^{\dag}\left( t^{\prime
}\right) \right\} \right\rangle ,
\end{equation}
where, after Fourier transformation, $\omega^{+}\equiv\omega+i\eta$.

We first consider the mixed Green's functions $G_{\boldsymbol{k}\sigma
,d\sigma }^{R}\left( t,t^{\prime }\right) $ and $G_{d\sigma ,\boldsymbol{k}%
\sigma }^{R}\left( t,t^{\prime }\right) $. Their EOMs read 
\begin{equation}
\left\{ 
\begin{array}{c}
i\partial _{t}G_{\boldsymbol{k}\sigma ,d\sigma }^{R}\left( t,t^{\prime
}\right) =-i\theta \left( t-t^{\prime }\right) \left\langle \left\{ \left[
c_{\boldsymbol{k}\sigma },H\right] \left( t\right) ,d_{\sigma }^{\dag
}\left( t^{\prime }\right) \right\} \right\rangle , \\[7pt] 
-i\partial _{t^{\prime }}G_{d\sigma ,\boldsymbol{k}\sigma }^{R}\left(
t,t^{\prime }\right) =-i\theta \left( t-t^{\prime }\right) \left\langle
\left\{ d_{\sigma }\left( t\right) ,\left[ c_{\boldsymbol{k}\sigma }^{\dag
},H\right] \left( t^{\prime }\right) \right\} \right\rangle .%
\end{array}%
\right.
\end{equation}%
Using $\left[ c_{\boldsymbol{k}\sigma },H\right] =V_{\boldsymbol{k}\sigma
}^{\ast }d_{\sigma }+\varepsilon _{\boldsymbol{k}\sigma }c_{\boldsymbol{k}%
\sigma }$ and $\left[ c_{\boldsymbol{k}\sigma }^{\dag },H\right] =-\left[ c_{%
\boldsymbol{k}\sigma },H\right] ^{\dag }$, and performing Fourier
transformation, we obtain 
\begin{equation}
\left\{ 
\begin{array}{c}
G_{\boldsymbol{k}\sigma ,d\sigma }^{R}\left( \omega \right) =g_{\boldsymbol{k%
}\sigma ,\boldsymbol{k}\sigma }^{R}\left( \omega \right) V_{\boldsymbol{k}%
\sigma }^{\ast }G_{d\sigma ,d\sigma }^{R}\left( \omega \right) , \\[7pt] 
G_{d\sigma ,\boldsymbol{k}\sigma }^{R}\left( \omega \right) =G_{d\sigma
,d\sigma }^{R}\left( \omega \right) V_{\boldsymbol{k}\sigma }g_{\boldsymbol{k%
}\sigma ,\boldsymbol{k}\sigma }^{R}\left( \omega \right) .%
\end{array}%
\right.  \label{eq:eom_mixed_gf}
\end{equation}%
Here, $g_{\boldsymbol{k}\sigma ,\boldsymbol{k}^{\prime }\sigma }^{R}\left(
\omega \right) =\delta _{\boldsymbol{k},\boldsymbol{k}^{\prime }}/\left(
\omega ^{+}-\varepsilon _{\boldsymbol{k}\sigma }\right) $ is the bare
substrate Green's function.

We next consider the impurity Green's function $G_{d\sigma ,d\sigma
}^{R}\left( t,t^{\prime }\right) $, its EOM is 
\begin{align}
& i\partial _{t}G_{d\sigma ,d\sigma }^{R}\left( t,t^{\prime }\right) =\delta
\left( t-t^{\prime }\right)  \notag \\
& -i\theta \left( t-t^{\prime }\right) \left\langle \left\{ \left[ d_{\sigma
},H\right] \left( t\right) ,d_{\sigma }^{\dag }\left( t^{\prime }\right)
\right\} \right\rangle .
\end{align}%
Using 
\begin{equation}
\left[ d_{\sigma },H\right] =\sum_{\boldsymbol{k}}V_{\boldsymbol{k}\sigma
}c_{\boldsymbol{k}\sigma }+\mathcal{E}_{\sigma }d_{\sigma },
\label{eq:eom_commutator_impurity}
\end{equation}%
together with Eq.~(\ref{eq:eom_mixed_gf}), and then Fourier transforming, we
obtain 
\begin{equation}
G_{d\sigma ,d\sigma }^{R}\left( \omega \right) =\frac{1}{\omega ^{+}-%
\mathcal{E}_{\sigma }-\Sigma _{\sigma }^{R}\left( \omega \right) }
\label{eq:eom_impurity_gf}
\end{equation}%
where the impurity self-energy is 
\begin{align}
\Sigma _{\sigma }^{R}\left( \omega \right) & =\sum_{\boldsymbol{k}}V_{%
\boldsymbol{k}\sigma }g_{\boldsymbol{k}\sigma ,\boldsymbol{k}\sigma
}^{R}\left( \omega \right) V_{\boldsymbol{k}\sigma }^{\ast }  \notag \\
& =\left\vert \mathcal{A}_{ai}\right\vert ^{2}g_{s\sigma }^{R}(\boldsymbol{r}%
_{d},\boldsymbol{r}_{d};\omega ).  \label{eq:eom_self_energy}
\end{align}

Finally, the EOM for $G_{\boldsymbol{k}\sigma ,\boldsymbol{k}^{\prime
}\sigma }^{R}\left( t,t^{\prime }\right) $ is 
\begin{align}
& i\partial _{t}G_{\boldsymbol{k}\sigma ,\boldsymbol{k}^{\prime }\sigma
}^{R}\left( t,t^{\prime }\right) =\delta \left( t-t^{\prime }\right) \delta
_{\boldsymbol{k},\boldsymbol{k}^{\prime }}  \notag \\
& -i\theta \left( t-t^{\prime }\right) \left\langle \left\{ \left[ c_{%
\boldsymbol{k}\sigma },H\right] \left( t\right) ,c_{\boldsymbol{k}^{\prime
}\sigma }^{\dag }\left( t^{\prime }\right) \right\} \right\rangle .
\end{align}%
After Fourier transformation, one obtains 
\begin{align}
& G_{\boldsymbol{k}\sigma ,\boldsymbol{k}^{\prime }\sigma }^{R}\left( \omega
\right) =g_{\boldsymbol{k}\sigma ,\boldsymbol{k}^{\prime }\sigma }^{R}\left(
\omega \right)  \notag \\
& +g_{\boldsymbol{k}\sigma ,\boldsymbol{k}\sigma }^{R}\left( \omega \right)
V_{\boldsymbol{k}\sigma }^{\ast }G_{d\sigma ,d\sigma }^{R}\left( \omega
\right) V_{\boldsymbol{k}^{\prime }\sigma }g_{\boldsymbol{k}^{\prime }\sigma
,\boldsymbol{k}^{\prime }\sigma }^{R}\left( \omega \right) .
\label{eq:eom_substrate_gf}
\end{align}

Equations~(\ref{eq:eom_mixed_gf}), (\ref{eq:eom_impurity_gf}), and (\ref%
{eq:eom_substrate_gf}) are the Green's-function identities used in the
derivation of the local spectral function below.

\section{Local spectral functions}

\label{app:local_spectral_function}

We now derive the local spectral function used in the main text. The
spin-resolved STM tunneling signal is proportional to the spectral weight
projected onto the tunneling channel, 
\begin{equation}
A_{\sigma}(d,\theta)=-2\func{Im}\left\langle \boldsymbol{r}_{t}
\sigma\right\vert \hat{V}_{t}\hat{G}^{R}(E_{F})\hat{V}_{t}^{\dagger}\left%
\vert \boldsymbol{r}_{t}\sigma\right\rangle .
\end{equation}
Here, $\hat{G}^{R}(\omega)=(\omega^{+}-H)^{-1}$ is the full retarded Green's
function of the coupled impurity-substrate system, excluding the STM tip.
The operator notation is related to the second-quantized representation in
the standard way \cite{BruusFlensberg2004}. The tip-system coupling is 
\begin{equation}
\hat{V}_{t}=\sum_{\sigma}\left( t_{d}\left\vert \boldsymbol{r}_{t}
\sigma\right\rangle \left\langle d\sigma\right\vert +\sum_{\boldsymbol{k}
}t_{\boldsymbol{k}\sigma}\left\vert \boldsymbol{r}_{t}\sigma\right\rangle
\left\langle \boldsymbol{k}\sigma\right\vert \right) +\mathrm{H.c.}
\end{equation}
where $\mathrm{H.c.}$ denotes the Hermitian conjugate. The corresponding
tunneling amplitudes are 
\begin{equation}
\left\{ 
\begin{array}{c}
\displaystyle t_{d}=\mathcal{A}_{it}\left\langle \boldsymbol{r}_{t} \sigma%
\middle|d\sigma\right\rangle , \\[6pt] 
\displaystyle t_{\boldsymbol{k}\sigma}=\mathcal{A}_{at}\left\langle 
\boldsymbol{r}_{t}\sigma\middle|\boldsymbol{k}\sigma\right\rangle .%
\end{array}
\right.  \label{eq:tip_tunneling_amplitudes}
\end{equation}

Using the completeness relation in the impurity-substrate Hilbert space, 
\begin{equation}
1=\sum_{\sigma}\left( \left\vert d\sigma\right\rangle \left\langle
d\sigma\right\vert +\sum_{\boldsymbol{k}}\left\vert \boldsymbol{k}
\sigma\right\rangle \left\langle \boldsymbol{k}\sigma\right\vert \right) ,
\end{equation}
the local spectral weight can be expanded as 
\begin{align}
& \left\langle \boldsymbol{r}_{t}\sigma\right\vert \hat{V}_{t}\hat{G}
^{R}(\omega)\hat{V}_{t}^{\dagger}\left\vert \boldsymbol{r}_{t}\sigma
\right\rangle  \notag \\
& =t_{d}G_{d\sigma,d\sigma}^{R}(\omega)t_{d}^{\ast}+\sum_{\boldsymbol{k} ,%
\boldsymbol{k}^{\prime}}t_{\boldsymbol{k}\sigma}G_{\boldsymbol{k} \sigma,%
\boldsymbol{k}^{\prime}\sigma}^{R}(\omega)t_{\boldsymbol{k}^{\prime
}\sigma}^{\ast}  \notag \\
& \quad+\sum_{\boldsymbol{k}}\left[ t_{\boldsymbol{k}\sigma} G_{\boldsymbol{k%
}\sigma,d\sigma}^{R}(\omega)t_{d}^{\ast}+t_{d}G_{d\sigma ,\boldsymbol{k}%
\sigma}^{R}(\omega)t_{\boldsymbol{k}\sigma}^{\ast}\right] ,
\label{eq:tip_spectral_expansion}
\end{align}
where 
\begin{equation}
G_{a,b}^{R}(\omega)=\left\langle a\right\vert \hat{G}^{R}(\omega)\left\vert
b\right\rangle .
\end{equation}
Substituting Eqs.~(\ref{eq:eom_mixed_gf}), (\ref{eq:eom_substrate_gf}), and (%
\ref{eq:tip_tunneling_amplitudes}) into Eq.~(\ref{eq:tip_spectral_expansion}%
), and identifying the momentum sums with real-space substrate Green's
functions, one obtains 
\begin{align}
& \left\langle \boldsymbol{r}_{t}\sigma\right\vert \hat{V}_{t}\hat{G}
^{R}(E_{F})\hat{V}_{t}^{\dagger}\left\vert \boldsymbol{r}_{t}\sigma
\right\rangle  \notag \\
& =\left\vert \mathcal{A}_{at}\right\vert ^{2}g_{s\sigma}^{R}(\boldsymbol{r}
_{t},\boldsymbol{r}_{t};E_{F})+\frac{N_{\sigma}}{\Delta_{\sigma}
-\Lambda+i\Gamma},
\end{align}
with 
\begin{align}
N_{\sigma} & =\left[ t_{d}+\mathcal{A}_{at}\mathcal{A}_{ai}^{\ast}g_{s%
\sigma} ^{R}(\boldsymbol{r}_{t},\boldsymbol{r}_{d};E_{F})\right]  \notag \\
& \times\left[ t_{d}^{\ast}+\mathcal{A}_{at}^{\ast}\mathcal{A}%
_{ai}g_{s\sigma}^{R}(\boldsymbol{r}_{d},\boldsymbol{r}_{t};E_{F})\right] .
\end{align}

For the contact-potential geometry considered here, the tunneling matrix
element $t_{d}$, which connects the tip position $\boldsymbol{r}_{t}$ to the
impurity orbital centered at $\boldsymbol{r}_{d}$, carries the same Peierls
phase as $g_{s\sigma}^{R}(\boldsymbol{r}_{t},\boldsymbol{r}_{d};E_{F})$ in
the presence of a magnetic field. Thus, the Peierls phases cancel between
the two brackets entering $N_{\sigma}$. This can be made explicit by writing 
\begin{equation}
t_{d}=e^{-i\chi(\boldsymbol{r}_{t},\boldsymbol{r}_{d})}\tilde{t}_{d}.
\end{equation}
For simplicity, we omit the Peierls phases of both $t_{d}$ and $g_{s\sigma
}^{R}(\boldsymbol{r}_{t},\boldsymbol{r}_{d};E_{F})$ in the following. Within
this convention, the remaining part of the substrate Green's function
satisfies 
\begin{equation}
g_{s\sigma}^{R}(\boldsymbol{r}_{t},\boldsymbol{r}_{d};E_{F})=g_{s\sigma}
^{R}(\boldsymbol{r}_{d},\boldsymbol{r}_{t};E_{F}).
\end{equation}

Taking $\mathcal{A}_{at}$, $\mathcal{A}_{ai}$, and $t_{d}$ to be real, and
using the dimensionless Green's function defined in the main text, the
numerator $N_{\sigma}$ can be written as 
\begin{equation}
N_{\sigma}=\left\{ \mathcal{A}_{at}\mathcal{A}_{ai}^{\ast}\func{Im}\left[
g_{s\sigma}^{R}(\boldsymbol{r}_{t},\boldsymbol{r}_{t};E_{F})\right] \right\}
^{2}Z_{\sigma}^{2},
\end{equation}
where 
\begin{equation}
Z_{\sigma}=\bar{g}_{s\sigma}^{R}(d,\theta)+\lambda
\end{equation}
with 
\begin{equation}
\lambda=-\frac{t_{d}}{\mathcal{A}_{at}\mathcal{A}_{ai}^{\ast}\func{Im}\left[
g_{s\sigma}^{R}(\boldsymbol{r}_{t},\boldsymbol{r}_{t};E_{F})\right] }.
\end{equation}
The dimensionless parameter $\lambda$ measures the direct tip-impurity
tunneling amplitude relative to the substrate-mediated tunneling scale. With
these definitions, the local spectral function can be cast into a
generalized Fano form, 
\begin{align}
& A_{\sigma}(d,\theta)=A_{0}(d,\theta)\times  \notag \\
& \left\{ 1+C_{\sigma}\left[ 1-\frac{\left[ (\Delta_{\sigma}-\Lambda)/%
\Gamma+q_{\sigma}\right] ^{2}}{(\Delta_{\sigma}-\Lambda)^{2}/\Gamma^{2}+1}%
\right] \right\} ,  \label{eq:sf_fano_tip}
\end{align}
where 
\begin{equation}
\left\{ 
\begin{array}{c}
\displaystyle C_{\sigma}=\left[ \func{Re}Z_{\sigma}\right] ^{2}, \\[10pt] 
\displaystyle q_{\sigma}=\tan\left[ \arg Z_{\sigma}\right] .%
\end{array}
\right.
\end{equation}

In the weak direct tip-impurity coupling limit, namely $\lambda\rightarrow0$%
, the direct tip-impurity contribution and the interference terms involving $%
t_{d}$ in Eq.~(\ref{eq:tip_spectral_expansion}) are suppressed. The local
spectral function is then dominated by the substrate-projected contribution, 
\begin{equation}
A_{\sigma}(d,\theta)\simeq-2\left\vert \mathcal{A}_{at}\right\vert ^{2}\func{%
Im}\left[ G_{s\sigma}^{R}(\boldsymbol{r}_{t},\boldsymbol{r}_{t};E_{F})\right]
,
\end{equation}
with%
\begin{align}
& G_{s\sigma}^{R}(\boldsymbol{r},\boldsymbol{r}^{\prime};\omega)=g_{s\sigma
}^{R}(\boldsymbol{r},\boldsymbol{r}^{\prime};\omega)  \notag \\
& \quad+|\mathcal{A}_{ai}|^{2}g_{s\sigma}^{R}(\boldsymbol{r},\boldsymbol{r}%
_{d};\omega)G_{d\sigma,d\sigma}^{R}(\omega)g_{s\sigma}^{R}(\boldsymbol{r}%
_{d},\boldsymbol{r}^{\prime};\omega)
\end{align}
the full retarded Green's function projected onto the substrate subspace.
Using the definitions above, Eq.~(\ref{eq:sf_fano_tip}) reduces to the local
spectral-function Fano form in this limit, where $C_{\sigma}$ and $%
q_{\sigma} $ reduce to 
\begin{equation}
\left\{ 
\begin{array}{c}
\displaystyle C_{\sigma}=\left\{ \func{Re}\left[ \bar{g}_{s\sigma
}^{R}(d,\theta)\right] \right\} ^{2}, \\[10pt] 
\displaystyle q_{\sigma}=\tan\left\{ \arg\left[ \bar{g}_{s\sigma}^{R}(d,%
\theta)\right] \right\} .%
\end{array}
\right.
\end{equation}

\section{Substrate Green's function under a perpendicular magnetic field}

\label{app:substrate_gf}

This appendix derives the bare substrate Green's function $%
g_{s\sigma}^{R}\left( \boldsymbol{r},\boldsymbol{r}^{\prime};\omega\right) $
used in the main text. The main technical difficulty comes from the
simultaneous presence of the vector potential and the altermagnetic cross
term, which is proportional to $k_{x}k_{y}$ in the absence of a magnetic
field. We treat this by rotating the coordinate system to eliminate the
cross term, imposing the Landau gauge in the rotated frame, evaluating the
Green's function there, and finally transforming the result back to the
original frame with the appropriate gauge compensation.

Following the path-integral approach of Ref.~\cite{Kleinert2009Path}, we
start from the two-point propagator $K_{\sigma}\left( \boldsymbol{R}%
_{b}t_{b},\boldsymbol{R}_{a}t_{a}\right) $. After evaluating this
propagator, we perform a Fourier transform to obtain the retarded substrate
Green's function $g_{s\sigma}^{R}\left( \boldsymbol{r};\boldsymbol{r}%
^{\prime};\omega\right) $. The Zeeman splitting is neglected in this
derivation, since it only shifts the energy argument and does not affect the
coordinate transformation or the path-integral evaluation.

We first rotate the coordinate axes clockwise by $\pi /4$: 
\begin{align}
\hat{h}_{s}\left( \boldsymbol{k},\sigma \right) & =\frac{\hbar ^{2}}{2m}%
\left( k_{x}^{2}+k_{y}^{2}+2\sigma \mathcal{J}k_{x}k_{y}\right)  \notag \\
& =\frac{\hbar ^{2}k_{X}^{2}}{2m_{X}}+\frac{\hbar ^{2}k_{Y}^{2}}{2m_{Y}},
\end{align}%
with $X/Y=\left( x\mp y\right) /\sqrt{2}$, $k_{X/Y}=\left( k_{x}\mp
k_{y}\right) /\sqrt{2}$, and $m_{X/Y}=m/\left( 1\mp \sigma \mathcal{J}%
\right) $. Here, capital letters denote the rotated frame. Imposing the
Landau gauge $A=BX\hat{e}_{Y}$ and applying the minimal substitution $%
p\rightarrow p+eA/c$, the Hamiltonian becomes 
\begin{equation}
\hat{h}_{s}=\frac{P_{X}^{2}}{2m_{X}}+\frac{P_{Y}^{2}}{2m_{Y}}+\omega
_{Y}XP_{Y}+\frac{1}{2}m_{Y}\omega _{Y}^{2}X^{2},
\label{eq:rotated_hamiltonian}
\end{equation}%
where $\omega _{Y}=eB/m_{Y}c$.

The propagator associated with Eq.~(\ref{eq:rotated_hamiltonian}), 
\begin{equation}
K_{\sigma }\left( \boldsymbol{R}_{b},t_{b};\boldsymbol{R}_{a},t_{a}\right)
=\left\langle \boldsymbol{R}_{b}\sigma \right\vert \exp \left[ -\frac{i}{%
\hbar }\left( t_{b}-t_{a}\right) \hat{h}_{s}\right] \left\vert \boldsymbol{R}%
_{a}\sigma \right\rangle ,
\end{equation}%
reduces, after integrating out the $Y$ degree of freedom, to that of a
one-dimensional harmonic oscillator. Since the intermediate steps are
standard, we present the result directly: 
\begin{widetext}
	\begin{equation}
		K_{\sigma}\left(  \boldsymbol{R}_{b},t_{b};\boldsymbol{R}_{a},t_{a}\right)
		=e^{-i\chi(\boldsymbol{R}_{b},\boldsymbol{R}_{a})}\frac{1}{\sqrt{1-\mathcal{J}^{2}}}\frac{m}{2\pi i\hbar}\frac
		{\omega_{L}/2}{\sin\frac{\omega_{L}\left(  t_{b}-t_{a}\right)  }{2}}%
		\exp\left\{  \frac{i}{\hbar}\frac{\omega_{L}}{4}\left[  m_{X}\left(
		X_{b}-X_{a}\right)  ^{2}+m_{Y}\left(  Y_{b}-Y_{a}\right)  ^{2}\right]
		\cot\frac{\omega_{L}\left(  t_{b}-t_{a}\right)  }{2}\right\}  .
		\label{eq:propagator_kernel}
	\end{equation}
\end{widetext}with $\omega _{L}=\sqrt{1-\mathcal{J}^{2}}\cdot eB/mc$, where 
\begin{eqnarray}
\chi \left( \boldsymbol{R}_{b},\boldsymbol{R}_{a}\right) &=&\frac{e}{\hbar c}%
\int_{\boldsymbol{R}_{a}}^{\boldsymbol{R}_{b}}\boldsymbol{A}(\boldsymbol{R}%
)\cdot d\boldsymbol{R}  \notag \\
&=&\frac{eB}{\hbar c}\frac{X_{a}+X_{b}}{2}\left( Y_{b}-Y_{a}\right)
\end{eqnarray}%
is the Peierls phase, which is independent of $t_{b}-t_{a}$.

Next, we perform the Fourier transformation: 
\begin{widetext}
\begin{align}
g_{s\sigma}^{R}\left(  \boldsymbol{R},0;\omega\right)   &  =-\frac{i}{\hbar
}\int_{-\infty}^{\infty}dt\exp\left(  \frac{i}{\hbar}E^{+}t\right)
\times\theta\left(  t\right)  K_{\sigma}\left(  \boldsymbol{R},t;0,0\right)
\nonumber\\
&  =-\frac{1}{\sqrt{1-\mathcal{J}^{2}}}\frac{m}{2\pi\hbar^{2}}\int_{0}%
^{\infty}dt\,\frac{\omega_{L}/2}{\sin\frac{\omega_{L}t}{2}}\exp\left\{
\frac{i}{\hbar}\left[  E^{+}t+\frac{\rho\omega_{L}}{4}\cot\frac{\omega_{L}%
t}{2}\right]  \right\}  , \label{eq:gf_time_integral}%
\end{align}
\end{widetext}where the Peierls phase is temporarily omitted, since it is
independent of time. Here $\boldsymbol{R}=(X,Y)$, $E^{+}=\hbar \omega ^{+}$,
and $\rho =m_{X}X^{2}+m_{Y}Y^{2}$. We now define $q=\exp \left( -i\omega
_{L}t\right) $, $z=\rho \omega _{L}/2\hbar $, and $a=1/2-E^{+}/\hbar \omega
_{L}$. Then $\sin \left( \omega _{L}t/2\right) =\left[ \exp \left( i\omega
_{L}t/2\right) -\exp \left( -i\omega _{L}t/2\right) \right] /2i=\left(
1-q\right) /2iq^{1/2}$, $\cot \left( \omega _{L}t/2\right) =i\left(
1+q\right) /\left( 1-q\right) $, and $\exp \left( iE^{+}t/\hbar \right)
=q^{-E^{+}/\hbar \omega _{L}}$. Thus, the integrand in Eq.~(\ref%
{eq:gf_time_integral}) can be rewritten as 
\begin{align}
& \frac{1}{\sin \frac{\omega _{L}t}{2}}\exp \left\{ \frac{i}{\hbar }\left[
E^{+}t+\frac{\rho \omega _{L}}{4}\cot \frac{\omega _{L}t}{2}\right] \right\}
\notag \\
& =2i\,e^{-z/2}\,q^{a}\frac{1}{1-q}\exp \left( -\frac{zq}{1-q}\right) .
\end{align}%
Next, by employing the generating function of the Laguerre polynomials, 
\begin{equation}
\sum_{n=0}^{\infty }L_{n}(z)q^{n}=\frac{1}{1-q}\exp \left( -\frac{zq}{1-q}%
\right) ,
\end{equation}%
we obtain 
\begin{equation}
q^{a}\frac{1}{1-q}\exp \left( -\frac{zq}{1-q}\right) =\sum_{n=0}^{\infty
}L_{n}(z)q^{n+a}.
\end{equation}%
Substituting this into Eq.~(\ref{eq:gf_time_integral}) gives 
\begin{align}
& g_{s\sigma }^{R}\left( \boldsymbol{R},0;\omega \right) =  \notag \\
& -\frac{1}{\sqrt{1-\mathcal{J}^{2}}}\frac{m}{2\pi \hbar ^{2}}%
e^{-z/2}\sum_{n=0}^{\infty }\frac{L_{n}(z)}{n+a}.
\end{align}%
Finally, using the identities 
\begin{equation}
\left\{ 
\begin{array}{c}
\sum_{n=0}^{\infty }\frac{L_{n}(z)}{n+a}=\Gamma _{\mathrm{E}}(a)\,U(a,1,z),
\\[10pt] 
W_{\kappa ,\mu }(z)=z^{\mu +1/2}e^{-z/2}U\left( \mu -\kappa +1/2,\,2\mu
+1,\,z\right) ,%
\end{array}%
\right.
\end{equation}%
with $\Gamma _{\mathrm{E}}(a)$ the Euler Gamma function, $U(a,b,z)$ the
Tricomi function, and $W_{\kappa ,\mu }(z)$ the Whittaker function, the
retarded Green's function in the frequency domain is obtained as 
\begin{align}
& g_{s\sigma }^{R}\left( \boldsymbol{R},0;\omega \right) =  \notag \\
& -\frac{1}{\sqrt{1-\mathcal{J}^{2}}}\frac{m}{2\pi \hbar ^{2}}\Gamma _{%
\mathrm{E}}\left( \frac{1}{2}-\zeta ^{+}\right) \frac{W_{\zeta
^{+},\,0}\left( R_{\sigma }^{2}\right) }{R_{\sigma }}.
\end{align}%
Here $\zeta ^{+}=\omega ^{+}/\omega _{L}$, $R_{\sigma }^{2}=z=\rho \omega
_{L}/2\hbar $, and $l_{B}=\sqrt{\hbar c/eB}$.

Finally, we compensate for the gauge difference and rotate back to the
original frame. First, from 
\begin{align}
\left( X,Y\right) & =\frac{1}{\sqrt{2}}\left( x-y,x+y\right)  \notag \\
& =r\left( \cos \left( \theta +\frac{\pi }{4}\right) ,\sin \left( \theta +%
\frac{\pi }{4}\right) \right) ,
\end{align}%
one finds that $R_{\sigma }^{2}=\left( \kappa _{\sigma }k_{F}d\right)
^{2}/4\zeta $ and $\kappa _{\sigma }=\sqrt{\left( 1-\sigma \mathcal{J}\sin
2\theta \right) /\left( 1-\mathcal{J}^{2}\right) }$, where we have
specialized to $\hbar \omega =E_{F}=\hbar ^{2}k_{F}^{2}/2m$. Second, for two
Green's functions defined in the gauges $\boldsymbol{A}$ and $\boldsymbol{A}%
^{\prime }=\boldsymbol{A}+\nabla \Lambda _{\mathrm{g}}$, the Peierls phase
changes by a boundary term, 
\begin{eqnarray}
&&\chi \left( \boldsymbol{R}_{b},\boldsymbol{R}_{a}\right) |_{\boldsymbol{A}%
^{\prime }}=\frac{e}{\hbar c}\int_{\boldsymbol{R}_{a}}^{\boldsymbol{R}_{b}}%
\boldsymbol{A}^{\prime }(\boldsymbol{R})\cdot d\boldsymbol{R}  \notag \\
&=&\frac{e}{\hbar c}\int_{\boldsymbol{R}_{a}}^{\boldsymbol{R}_{b}}\left[ 
\boldsymbol{A}(\boldsymbol{R})+\nabla \Lambda _{\mathrm{g}}\left( 
\boldsymbol{R}\right) \right] \cdot d\boldsymbol{R}  \notag \\
&=&\chi \left( \boldsymbol{R}_{b},\boldsymbol{R}_{a}\right) |_{\boldsymbol{A}%
}+\frac{e}{\hbar c}\left[ \Lambda _{\mathrm{g}}\left( \boldsymbol{R}%
_{b}\right) -\Lambda _{\mathrm{g}}\left( \boldsymbol{R}_{a}\right) \right] .
\end{eqnarray}%
Using $\boldsymbol{A}^{\prime }=Bx\hat{\boldsymbol{e}}_{y}$ and $\boldsymbol{%
A}=BX\hat{\boldsymbol{e}}_{Y}$, one obtains 
\begin{equation}
\Lambda _{\mathrm{g}}=\frac{B}{2}\left( -\frac{1}{2}X^{2}-XY+\frac{1}{2}%
Y^{2}\right) .
\end{equation}%
The corresponding Peierls factor in the original gauge is 
\begin{equation}
e^{-i\chi \left( \boldsymbol{R}_{b},\boldsymbol{R}_{a}\right) |_{\boldsymbol{%
A}}}=\exp \left[ -i\frac{eB}{2\hbar c}\left( x_{a}+x_{b}\right) \left(
y_{b}-y_{a}\right) \right] .
\end{equation}%
In sum, the bare substrate Green's function in the Landau-quantized regime
is 
\begin{align}
& g_{s\sigma }^{R}\left( \boldsymbol{r}_{t},\boldsymbol{r}_{d};E_{F}\right)
=-\frac{1}{\sqrt{1-\mathcal{J}^{2}}}\frac{m}{2\pi \hbar ^{2}}\times  \notag
\\
& e^{-i\chi }\Gamma _{\mathrm{E}}\left( \frac{1}{2}-\zeta ^{+}\right) \frac{%
W_{\zeta ^{+},0}\left( R_{\sigma }^{2}\right) }{R_{\sigma }}
\label{eq:landau_substrate_gf}
\end{align}%
with 
\begin{equation}
\chi \left( \boldsymbol{r}_{t},\boldsymbol{r}_{d}\right) =\frac{e B}{2\hbar c%
}\left( x_{d}+x_{t}\right) \left( y_{t}-y_{d}\right).
\end{equation}%
The zero-field-regime expression is recovered through a two-step limiting
procedure. First, the Landau-level structure is smeared out by taking $\hbar
\omega _{L}\rightarrow \eta $. Then, in the further limit $\eta \rightarrow
0^{+}$, one obtains 
\begin{align}
& \Gamma _{\mathrm{E}}\left( \frac{1}{2}-\zeta ^{+}\right) \frac{W_{\zeta
^{+},0}\left( R_{\sigma }^{2}\right) }{R_{\sigma }}\rightarrow  \notag \\
& i\pi H_{0}^{(1)}\left( 2\sqrt{\zeta R_{\sigma }^{2}}\right) =i\pi
H_{0}^{(1)}\left( \kappa _{\sigma }k_{F}d\right) .
\label{eq:landau_to_zero-field_limit}
\end{align}%
Consequently, we have%
\begin{equation}
g_{s\sigma }^{R}\left( \boldsymbol{r}_{t},\boldsymbol{r}_{d};E_{F}\right) =-%
\frac{1}{\sqrt{1-\mathcal{J}^{2}}}\frac{im}{2\hbar ^{2}}H_{0}^{(1)}\left(
\kappa _{\sigma }k_{F}d\right) .
\end{equation}%
In other words, the term \textquotedblleft Landau-quantized
regime\textquotedblright\ refers to the regime in which the Landau-level
spacing exceeds the effective broadening, so that the Landau-level-induced
structure is resolved. It does not necessarily imply the quantum-limit
regime.

\section{Renormalization of the impurity Green's function}

\label{app:impurity_renormalization}

This appendix explains the regularization of the impurity retarded
self-energy in Eq.~(\ref{eq:impurity_self_energy}). The regularization is
needed because the coincident-point Green's function $g_{s\sigma}^{R}(%
\boldsymbol{r}_{d},\boldsymbol{r}_{d};\omega)$ is logarithmically divergent
in two dimensions. This ultraviolet divergence reflects the mismatch between
the contact-potential approximation for the impurity-substrate coupling in
Eq.~(\ref{eq:hyb_matrix_element}), which probes short-range physics, and the
zero-field description of the substrate, which is valid only at long
wavelengths. We regularize this divergence by introducing a short-distance
cutoff $a$, which may be identified with the lattice scale or the spatial
extent of the impurity orbital. Equivalently, we evaluate $g_{s\sigma}^{R}(%
\boldsymbol{r},\boldsymbol{r}_{d};\omega)$ at $|\boldsymbol{r}-\boldsymbol{r}%
_{d}|=a$ instead of taking the formal limit $\boldsymbol{r}\rightarrow%
\boldsymbol{r}_{d}$.

We first consider the zero-field regime in the absence of Zeeman splitting.
Using the short-distance asymptotic form 
\begin{equation}
\lim_{z\rightarrow0}H_{0}^{(1)}(z) =1+\frac{2i}{\pi}\left( \gamma+\ln\frac {z%
}{2}\right) ,
\end{equation}
where $\gamma$ is the Euler constant, we obtain 
\begin{align}
& \lim_{a\rightarrow0}g_{s\sigma}^{R}\left( \boldsymbol{a},0;E_{F}\right) = 
\notag \\
& \frac{1}{\sqrt{1-\mathcal{J}^{2}}} \left[ \frac{m}{\pi\hbar^{2}} \left(
\gamma+\ln\frac{\kappa_{\sigma}k_{F}a}{2}\right) -\frac{im}{2\hbar^{2}} %
\right] .
\end{align}
Since the coincident-point regularization should not depend on the direction
from which the short-distance limit is approached, the angular dependence
through $\kappa_{\sigma}(\theta)$ is an artifact of applying the continuum
Green's function down to the cutoff scale. We remove this unphysical angular
dependence by replacing $\ln\kappa_{\sigma}$ with its angular average and
defining $\bar{\kappa}$ by 
\begin{equation}
\frac{1}{2\pi}\int_{0}^{2\pi}d\theta\,\ln\kappa_{\uparrow}(\theta) = \frac {1%
}{2\pi}\int_{0}^{2\pi}d\theta\,\ln\kappa_{\downarrow}(\theta) \equiv\ln \bar{%
\kappa}.
\end{equation}
The equality between the spin-up and spin-down angular averages follows from
the $\hat{C}_{4}\hat{T}$ symmetry. Thus, 
\begin{align}
& \lim_{a\rightarrow0}g_{s\sigma}^{R}\left( \boldsymbol{a},0;E_{F}\right)
\rightarrow  \notag \\
& \frac{1}{\sqrt{1-\mathcal{J}^{2}}} \left[ \frac{m}{\pi\hbar^{2}} \left(
\gamma+\ln\frac{\bar{\kappa}k_{F}a}{2}\right) -\frac{im}{2\hbar^{2}} \right]
.
\end{align}

The renormalization condition is usually imposed by absorbing the
ultraviolet-divergent real part of the impurity self-energy into the bare
impurity level. Here, the finite part of the real self-energy is also
included in the same definition of the renormalized impurity level for
brevity. With this convention, at $\omega=E_{F}$, the renormalized impurity
level is denoted by $\xi$, namely 
\begin{equation}
\xi=\varepsilon_{0} +\frac{1}{\sqrt{1-\mathcal{J}^{2}}} \frac{m|\mathcal{A}
_{ai}|^{2}}{\pi\hbar^{2}} \left( \ln\frac{\bar{\kappa}k_{F}a}{2}
+\gamma\right) .  \label{eq:renormalized_impurity_level}
\end{equation}
Accordingly, in the zero-field regime and in the absence of Zeeman
splitting, the impurity Green's function becomes 
\begin{equation}
G_{d\sigma,d\sigma}^{R}(E_{F})=\frac{1}{\Delta+i\Gamma},
\end{equation}
where $\Delta=E_{F}-\xi$ is the impurity-level detuning, and 
\begin{equation}
\Gamma= \frac{1}{\sqrt{1-\mathcal{J}^{2}}} \frac{m|\mathcal{A}_{ai}|^{2}}{
2\hbar^{2}}
\end{equation}
is the impurity linewidth.

We now include the impurity and substrate Zeeman splittings, denoted by $%
\varepsilon_{d}$ and $\varepsilon_{s}$, respectively. In the zero-field
regime, the substrate Zeeman splitting shifts the spin-dependent Fermi wave
vector to 
\begin{equation}
k_{F}^{\sigma}=\sqrt{\frac{2m(E_{F}-\sigma\varepsilon_{s})}{\hbar^{2}}},
\end{equation}
where $E_{F}-\sigma\varepsilon_{s}>0$ is assumed. The regularized
coincident-point Green's function then becomes 
\begin{align}
& \lim_{a\rightarrow0}g_{s\sigma}^{R}\left( \boldsymbol{a},0;E_{F}\right)
\rightarrow  \notag \\
& \frac{1}{\sqrt{1-\mathcal{J}^{2}}}\left[ \frac{m}{\pi\hbar^{2}}\left(
\gamma+\ln\frac{\bar{\kappa}k_{F}a}{2}+\ln\frac{k_{F}^{\sigma}}{k_{F}}%
\right) -\frac{im}{2\hbar^{2}}\right] .
\end{align}
Using the same renormalization condition as in Eq.~(\ref%
{eq:renormalized_impurity_level}), we obtain 
\begin{equation}
G_{d\sigma,d\sigma}^{R}(E_{F})=\frac{1}{\Gamma}\left[ \dfrac{\Delta
-\sigma\varepsilon_{d}}{\Gamma}-\dfrac{1}{\pi}\ln\left( 1-\sigma \frac{%
\varepsilon_{s}}{E_{F}}\right) +i\right] ^{-1}.
\end{equation}
This expression shows the different roles played by the two Zeeman terms.
The substrate Zeeman splitting enters through the substrate energy scale and
gives a correction controlled by $\varepsilon_{s}/E_{F}$. By contrast, the
impurity Zeeman splitting directly shifts the resonant impurity level and is
measured relative to the much narrower linewidth $\Gamma$. Thus, in the
zero-field-regime discussion of the main text, we retain the impurity Zeeman
splitting and neglect the substrate Zeeman splitting, which gives 
\begin{equation}
G_{d\sigma,d\sigma}^{R}(E_{F})=\frac{1}{\Delta_{\sigma}+i\Gamma}
\end{equation}
with $\Delta_{\sigma}=E_{F}-\left( \xi+\sigma\varepsilon_{d}\right) $ the
spin-dependent detuning.

We next consider the Landau-quantized regime. Using the short-distance
asymptotic form 
\begin{equation}
\lim_{z\rightarrow 0}W_{\lambda ,0}(z)=-\frac{\sqrt{z}}{\Gamma _{\mathrm{E}%
}\left( 1/2-\lambda \right) }\left[ \ln z+\psi \left( 1/2-\lambda \right)
+2\gamma \right] ,
\end{equation}%
and replacing $\ln \kappa _{\sigma }$ by the angular average $\ln \bar{\kappa%
}$, the regularized coincident-point Green's function at $\omega =E_{F}$ can
be written as 
\begin{align}
& g_{s\sigma }^{R}\left( \boldsymbol{a},0;E_{F}\right) =\frac{1}{\sqrt{1-%
\mathcal{J}^{2}}}\left\{ \frac{m}{\pi \hbar ^{2}}\left( \ln \frac{\bar{\kappa%
}k_{F}a}{2}+\gamma \right) \right.  \notag \\
& \left. +\frac{m}{2\pi \hbar ^{2}}\left[ -\ln \zeta +\psi \left( 1/2-\zeta
_{\sigma }^{+}\right) \right] \right\} .
\end{align}%
Here $\psi(z)$ is the digamma function, $\zeta _{\sigma }^{+}=\left(
E_{F}-\sigma \varepsilon _{s}+i\eta \right) /\hbar \omega _{L}$. The
ultraviolet divergence is contained in the same logarithmic short-distance
term as in the zero-field regime, and hence the same renormalization
condition, Eq.~(\ref{eq:renormalized_impurity_level}), can be used. The
remaining finite term gives the field-dependent correction to the impurity
self-energy. Including both impurity and substrate Zeeman splittings, one
obtains 
\begin{equation}
G_{d\sigma ,d\sigma }^{R}(E_{F})=\frac{1}{\Delta _{\sigma }-\Lambda _{\sigma
}+i\Gamma _{\sigma }},
\end{equation}%
with 
\begin{equation}
\left\{ 
\begin{array}{c}
\displaystyle\Lambda _{\sigma }=\frac{1}{\sqrt{1-\mathcal{J}^{2}}}\frac{m|%
\mathcal{A}_{ai}|^{2}}{2\pi \hbar ^{2}}\left\{ -\ln \zeta +\func{Re}\left[
\psi \left( 1/2-\zeta _{\sigma }^{+}\right) \right] \right\} , \\[8pt] 
\displaystyle\Gamma _{\sigma }=-\frac{1}{\sqrt{1-\mathcal{J}^{2}}}\frac{m|%
\mathcal{A}_{ai}|^{2}}{2\pi \hbar ^{2}}\func{Im}\left[ \psi \left( 1/2-\zeta
_{\sigma }^{+}\right) \right] .%
\end{array}%
\right.
\end{equation}

\begin{figure}[ptb]
\centering
\includegraphics[width=1\linewidth]{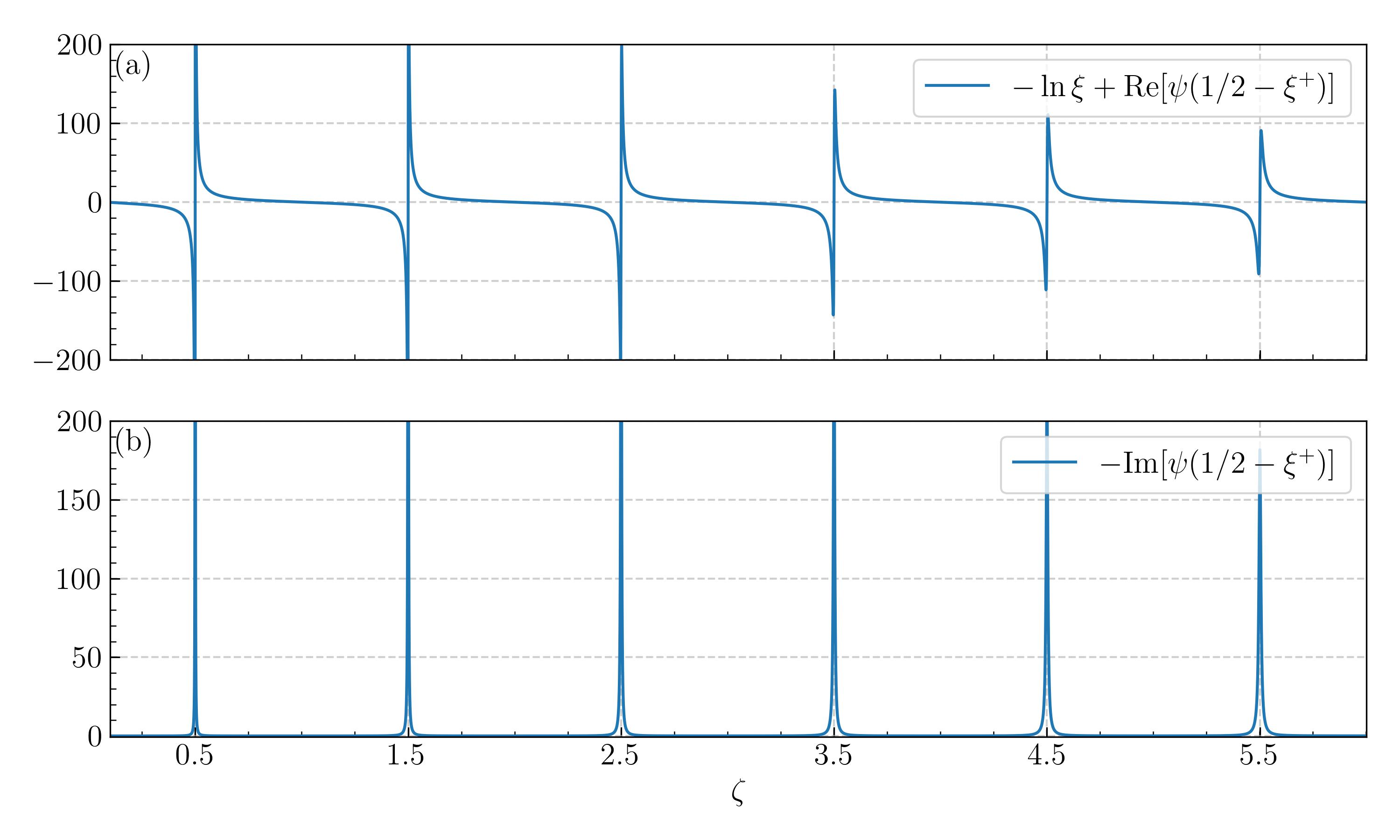}
\caption{Dependence of (a) the energy shift and (b) the linewidth on the
Landau-level filling factor $\protect\zeta$. In all panels, the broadening
is chosen as $\protect\eta=10^{-3}E_{F}$.}
\label{fig:figs1}
\end{figure}

Away from Landau-level resonances, the substrate Zeeman splitting gives only
a smooth spin-dependent correction to both the energy shift and the
linewidth (see Fig.~\ref{fig:figs1}). Since the regime in which the Fermi
level lies extremely close to a Landau-level resonance is not the focus of
the present work, we neglect the substrate Zeeman splitting while retaining
the impurity Zeeman splitting in the main text. This gives 
\begin{equation}
G_{d\sigma ,d\sigma }^{R}(E_{F})=\frac{1}{\Delta _{\sigma }-\Lambda +i\Gamma 
},
\end{equation}%
where 
\begin{equation}
\left\{ 
\begin{array}{c}
\displaystyle\Lambda =\frac{1}{\sqrt{1-\mathcal{J}^{2}}}\frac{m|\mathcal{A}%
_{ai}|^{2}}{2\pi \hbar ^{2}}\left\{ -\ln \zeta +\func{Re}\left[ \psi \left(
1/2-\zeta ^{+}\right) \right] \right\} , \\[8pt] 
\displaystyle\Gamma =-\frac{1}{\sqrt{1-\mathcal{J}^{2}}}\frac{m|\mathcal{A}%
_{ai}|^{2}}{2\pi \hbar ^{2}}\func{Im}\left[ \psi \left( 1/2-\zeta
^{+}\right) \right] ,%
\end{array}%
\right.
\end{equation}%
with $\zeta ^{+}=\left( E_{F}+i\eta \right) /\hbar \omega _{L}$.

Finally, the consistency between the zero-field and Landau-quantized
expressions can be seen by taking the successive limits $\hbar\omega
_{L}\rightarrow0$ and $\eta\rightarrow0^{+}$. In this limit, one has $%
\psi\left( 1/2-\zeta^{+}\right) \rightarrow\ln\zeta-i\pi$. Substituting this
asymptotic form into the Landau-quantized expression gives $\Lambda
\rightarrow0$ and 
\begin{equation}
\Gamma\rightarrow\frac{1}{\sqrt{1-\mathcal{J}^{2}}}\frac{m|\mathcal{A}%
_{ai}|^{2}}{2\hbar^{2}},
\end{equation}
thereby recovering the zero-field-regime result.

\section{Origin of the local spectral-function line-shape asymmetry}

\label{app:lse_asymmetry}

The Fano-type line-shape asymmetry in the local spectral function can be
understood most directly from the scattering-state representation. Although
often ignored for simplicity, such an asymmetry is in fact generic. The
essential reason is that a resonant amplitude is usually combined with other
complex amplitudes before the physical observable is obtained. For example,
the resonant Green's function contains both an even Lorentzian part and an
odd dispersive part, 
\begin{equation}
\frac{1}{\xi+i}=\frac{\xi}{\xi^{2}+1}-i\frac{1}{\xi^{2}+1},
\end{equation}
with $\xi$ a dimensionless detuning. When this resonant factor is multiplied
by a complex nonresonant amplitude and one takes either the imaginary part,
as in the local spectral function, or the modulus squared, as in an
intensity-like response, the even and odd components are generally mixed.
The resulting line shape therefore contains contributions from both parts.
This coexistence is precisely the structure of a Fano-type profile, since 
\begin{equation}
\frac{\left( \xi+q\right) ^{2}}{\xi^{2}+1} = 1+ \underbrace{ \frac{q^{2}-1}{%
\xi^{2}+1} }_{\mathrm{even}} + \underbrace{ \frac{2q\xi}{\xi^{2}+1} }_{%
\mathrm{odd}} .
\end{equation}
Thus, a symmetric Lorentzian line shape is not the generic case but a
special limit in which the odd dispersive contribution is absent or
suppressed.

When the direct tip-impurity tunneling channel is included, as in Eq.~(\ref%
{eq:sf_fano_tip}), the line shape asymmetry is usually attributed to the
off-diagonal Green's functions $G_{\boldsymbol{k}\sigma,d\sigma}^{R}$ and $%
G_{d\sigma,\boldsymbol{k}\sigma}^{R}$ [see Eq.~(\ref{eq:eom_mixed_gf})].
These terms are commonly referred to as interference terms in STM studies of
Fano line shapes, particularly for Kondo or adsorbate resonances \cite%
{Ujsaghy2000FanoSTM,plihal2001nonequilibrium,morr2017theory}, although the
local spectral function is obtained from the imaginary part of the Green's
function rather than from the modulus squared of a scattering amplitude. In
many simplified treatments, the tunneling amplitudes, such as $t_{d}$, $t_{%
\boldsymbol{k}\sigma}$, and $V_{\boldsymbol{k}\sigma}$, are taken to be
real. Similarly, in the present model one may take $t_{d}$, $\mathcal{A}%
_{at} $, and $\mathcal{A}_{ai}$ to be real and further apply the wide-band
approximation by neglecting the real part of $g_{s\sigma}^{R}(\boldsymbol{r}%
_{t},\boldsymbol{r}_{d};E_{F})$. In such cases, the $q$ factor reduces to,
in the phase convention where the Peierls phase is omitted, 
\begin{equation}
q_{\sigma}\simeq\frac{\mathcal{A}_{at}\mathcal{A}_{ai}^{\ast}\func{Im}\left[
g_{s\sigma}^{R}(\boldsymbol{r}_{t},\boldsymbol{r}_{d};E_{F})\right] }{t_{d}}.
\end{equation}
This expression has a simple physical interpretation: the Fano parameter is
essentially the ratio between the substrate-mediated indirect tunneling
amplitude and the direct tip-impurity tunneling amplitude. Beyond this
simplified picture, however, the line shape asymmetry is also influenced by
the complex phases of the tunneling amplitudes and of the substrate Green's
function. In the more general case, the Fano parameter should therefore be
understood as a phase-sensitive quantity rather than merely as a ratio of
two tunneling strengths.

However, in the absence of the direct tip-impurity tunneling channel, as
obtained from the $\lambda\rightarrow0$ limit of Eq.~(\ref%
{eq:local_fano_general}), the origin of the line shape asymmetry is less
transparent from the Green's-function expression alone, because it cannot be
simply attributed to the off-diagonal Green's functions. In this case, the
wave-function-based scattering theory, namely the Lippmann-Schwinger
equation (LSE), offers a useful complementary explanation.

The LSE starts from the inhomogeneous Schr\"{o}dinger equation, 
\begin{equation}
\left\{ 
\begin{array}{c}
\left( \varepsilon _{\boldsymbol{k}\sigma }^{+}-H_{0}\right) \left\vert \psi
_{\boldsymbol{k}\sigma }^{(+)}\right\rangle =V\left\vert \psi _{\boldsymbol{k%
}\sigma }^{(+)}\right\rangle , \\[6pt] 
\left\vert \psi _{\boldsymbol{k}\sigma }^{(+)}\right\rangle =\left\vert 
\boldsymbol{k}\sigma \right\rangle +\dfrac{1}{\varepsilon _{\boldsymbol{k}%
\sigma }^{+}-H_{0}}V\left\vert \psi _{\boldsymbol{k}\sigma
}^{(+)}\right\rangle ,%
\end{array}%
\right.
\end{equation}%
with $\varepsilon _{\boldsymbol{k}\sigma }^{+}=\varepsilon _{\boldsymbol{k}%
\sigma }+i\eta $. For the present model, there is a one-to-one
correspondence between a substrate state and the corresponding scattering
state, $\left\vert \boldsymbol{k}\sigma \right\rangle \rightarrow \left\vert
\psi _{\boldsymbol{k}\sigma }^{(+)}\right\rangle $, as formally guaranteed
by the M{\o }ller operator \cite{taylor2012scattering}. The scattering state
can be decomposed into an incoming component and a scattered component, 
\begin{equation}
\left\vert \psi _{\boldsymbol{k}\sigma }^{(+)}\right\rangle =\left\vert \psi
_{\boldsymbol{k}\sigma }^{(\mathrm{in})}\right\rangle +\left\vert \psi _{%
\boldsymbol{k}\sigma }^{(\mathrm{sc})}\right\rangle ,
\end{equation}%
where $\left\vert \psi _{\boldsymbol{k}\sigma }^{(\mathrm{in})}\right\rangle
=\left\vert \boldsymbol{k}\sigma \right\rangle $, and 
\begin{align}
& \left\vert \psi _{\boldsymbol{k}\sigma }^{(\mathrm{sc})}\right\rangle =V_{%
\boldsymbol{k}\sigma }G_{d\sigma ,d\sigma }^{R}\left( \varepsilon _{%
\boldsymbol{k}\sigma }\right) \left\vert d\sigma \right\rangle  \notag \\
\quad & +\sum_{\boldsymbol{q}}V_{\boldsymbol{k}\sigma }G_{d\sigma ,d\sigma
}^{R}\left( \varepsilon _{\boldsymbol{k}\sigma }\right) V_{\boldsymbol{q}%
\sigma }^{\ast }g_{\boldsymbol{q}\sigma }^{R}\left( \varepsilon _{%
\boldsymbol{k}\sigma }\right) \left\vert \boldsymbol{q}\sigma \right\rangle .
\end{align}%
For the present problem, one important advantage of the scattering-state
description is that it provides a complete orthogonal basis,%
\begin{equation}
\sum_{\boldsymbol{k}\sigma }\left\vert \psi _{\boldsymbol{k}\sigma
}^{(+)}\right\rangle \left\langle \psi _{\boldsymbol{k}\sigma
}^{(+)}\right\vert =\sum_{\sigma }\left( \left\vert d\sigma \right\rangle
\left\langle d\sigma \right\vert +\sum_{\boldsymbol{k}}\left\vert 
\boldsymbol{k}\sigma \right\rangle \left\langle \boldsymbol{k}\sigma
\right\vert \right) .
\end{equation}%
Here, the orthogonality is formally guaranteed by the M{\o }ller operator,
whereas the completeness can be checked through the spectral representation

\begin{equation}
\hat{G}^{R}\left( \omega\right) = \sum_{\boldsymbol{k}\sigma} \frac {%
\left\vert \psi_{\boldsymbol{k}\sigma}^{(+)}\right\rangle \left\langle \psi_{%
\boldsymbol{k}\sigma}^{(+)}\right\vert }{ \omega^{+}-\varepsilon _{%
\boldsymbol{k}\sigma} },
\end{equation}
which correctly reproduces the full retarded Green's function established
above.

Using this spectral representation and projecting onto the substrate
subspace, the local spectral function can be written as 
\begin{widetext}
	\begin{equation}
		A_{\sigma}\left(d,\theta\right)
		=
		-2\left\vert \mathcal{A}_{at}\right\vert ^{2}\operatorname{Im}
		\left[
		\sum_{\boldsymbol{k},\boldsymbol{k}^{\prime},\boldsymbol{q}}
		\left\langle
		\boldsymbol{r}_{t}\sigma
		\middle|
		\boldsymbol{k}\sigma
		\right\rangle
		\frac{
			\left\langle
			\boldsymbol{k}\sigma
			\middle|
			\psi_{\boldsymbol{q}\sigma}^{(+)}
			\right\rangle
			\left\langle
			\psi_{\boldsymbol{q}\sigma}^{(+)}
			\middle|
			\boldsymbol{k}^{\prime}\sigma
			\right\rangle
		}{
			E_{F}^{+}
			-
			\varepsilon_{\boldsymbol{q}\sigma}
		}
		\left\langle
		\boldsymbol{k}^{\prime}\sigma
		\middle|
		\boldsymbol{r}_{t}\sigma
		\right\rangle
		\right]
		=
		2\pi\left\vert \mathcal{A}_{at}\right\vert ^{2}\sum_{\boldsymbol{q}}
		\left|
		\left\langle
		\boldsymbol{r}_{t}\sigma
		\middle|
		\phi_{\boldsymbol{q}\sigma}^{(+)}
		\right\rangle
		\right|^{2}
		\delta
		\left(
		E_{F}
		-
		\varepsilon_{\boldsymbol{q}\sigma}
		\right).
	\label{eq:sf_substrate_spectral_representation}%
	\end{equation}
\end{widetext}Here, 
\begin{equation}
\left\vert \phi _{\boldsymbol{q}\sigma }^{(+)}\right\rangle =\sum_{%
\boldsymbol{k}}\left\vert \boldsymbol{k}\sigma \right\rangle \left\langle 
\boldsymbol{k}\sigma \middle|\psi _{\boldsymbol{q}\sigma
}^{(+)}\right\rangle =\left\vert \phi _{\boldsymbol{q}\sigma }^{(\mathrm{in}%
)}\right\rangle +\left\vert \phi _{\boldsymbol{q}\sigma }^{(\mathrm{sc}%
)}\right\rangle ,
\end{equation}%
stands for the scattering state projected onto the substrate subspace, with 
\begin{equation}
\left\{ 
\begin{array}{c}
\left\vert \phi _{\boldsymbol{q}\sigma }^{(\mathrm{in})}\right\rangle
=\left\vert \boldsymbol{q}\sigma \right\rangle , \\[6pt] 
\left\vert \phi _{\boldsymbol{q}\sigma }^{(\mathrm{sc})}\right\rangle =\sum_{%
\boldsymbol{k}}V_{\boldsymbol{q}\sigma }G_{d\sigma ,d\sigma }^{R}\left(
\varepsilon _{\boldsymbol{q}\sigma }\right) V_{\boldsymbol{k}\sigma }^{\ast
}g_{\boldsymbol{k}\sigma }^{R}\left( \varepsilon _{\boldsymbol{q}\sigma
}\right) \left\vert \boldsymbol{k}\sigma \right\rangle .%
\end{array}%
\right.
\end{equation}%
From this form it follows that the local spectral function can be decomposed
into three parts, 
\begin{equation}
A_{\sigma }\left( d,\theta \right) =A_{\sigma }^{\mathrm{inc}}\left(
d,\theta \right) +A_{\sigma }^{\mathrm{sc}}\left( d,\theta \right)
+A_{\sigma }^{\mathrm{int}}\left( d,\theta \right) ,
\end{equation}%
namely the incoming contribution, the scattered contribution, and the
interference term between them. Explicitly, 
\begin{align}
& \sum_{\boldsymbol{q}}\left\vert \left\langle \boldsymbol{r}_{t}\sigma %
\middle|\phi _{\boldsymbol{q}\sigma }^{(+)}\right\rangle \right\vert
^{2}\delta \left( E_{F}-\varepsilon _{\boldsymbol{q}\sigma }\right)  \notag
\\
& =\sum_{\boldsymbol{q}}\left\vert \left\langle \boldsymbol{r}_{t}\sigma %
\middle|\phi _{\boldsymbol{q}\sigma }^{(\mathrm{in})}\right\rangle
\right\vert ^{2}\delta \left( E_{F}-\varepsilon _{\boldsymbol{q}\sigma
}\right)  \notag \\
& \quad +\sum_{\boldsymbol{q}}\left\vert \left\langle \boldsymbol{r}%
_{t}\sigma \middle|\phi _{\boldsymbol{q}\sigma }^{(\mathrm{sc}%
)}\right\rangle \right\vert ^{2}\delta \left( E_{F}-\varepsilon _{%
\boldsymbol{q}\sigma }\right)  \notag \\
& \quad +\sum_{\boldsymbol{q}}\Big[\left\langle \phi _{\boldsymbol{q}\sigma
}^{(\mathrm{sc})}\middle|\boldsymbol{r}_{t}\sigma \right\rangle \left\langle 
\boldsymbol{r}_{t}\sigma \middle|\phi _{\boldsymbol{q}\sigma }^{(\mathrm{in}%
)}\right\rangle  \notag \\
& \qquad +\left\langle \phi _{\boldsymbol{q}\sigma }^{(\mathrm{in})}\middle|%
\boldsymbol{r}_{t}\sigma \right\rangle \left\langle \boldsymbol{r}_{t}\sigma %
\middle|\phi _{\boldsymbol{q}\sigma }^{(\mathrm{sc})}\right\rangle \Big]%
\delta \left( E_{F}-\varepsilon _{\boldsymbol{q}\sigma }\right) .
\end{align}

Substituting the explicit forms of $\left\vert \phi_{\boldsymbol{q}\sigma
}^{(\mathrm{in})}\right\rangle $ and $\left\vert \phi_{\boldsymbol{q}\sigma
}^{(\mathrm{sc})}\right\rangle $, one obtains 
\begin{equation}
\left\{ 
\begin{array}{c}
\displaystyle A_{\sigma}^{\mathrm{inc}}\left( d,\theta\right) =A_{0}\left(
d,\theta\right) , \\[12pt] 
\displaystyle A_{\sigma}^{\mathrm{sc}}\left( d,\theta\right) =\left\vert 
\mathcal{A}_{ai}\right\vert ^{4}A_{0}\left( d,\theta\right) \frac{\left\vert
g_{s\sigma}^{R}\left( \boldsymbol{r}_{t},\boldsymbol{r}_{d};\omega\right)
\right\vert ^{2}}{\left[ \Delta_{\sigma}-\Lambda\left( \omega\right) \right]
^{2}+\Gamma^{2}\left( \omega\right) }, \\[14pt] 
\displaystyle A_{\sigma}^{\mathrm{int}}\left( d,\theta\right) =2\left\vert 
\mathcal{A}_{at}\right\vert ^{2}\left\vert \mathcal{A}_{ai}\right\vert ^{2}%
\func{Im}\left\{ \frac{g_{s\sigma}^{R}\left( \boldsymbol{r}_{t},\boldsymbol{r%
}_{d};\omega\right) }{\left[ \Delta_{\sigma}-\Lambda\left( \omega\right) %
\right] +i\Gamma\left( \omega\right) }\right. \\[12pt] 
\displaystyle\left. \times\left[ g_{s\sigma}^{A}\left( \boldsymbol{r}_{d},%
\boldsymbol{r}_{t};\omega\right) -g_{s\sigma}^{R}\left( \boldsymbol{r}_{d},%
\boldsymbol{r}_{t};\omega\right) \right] \right\} .%
\end{array}
\right.
\end{equation}
Here, $g_{s\sigma}^{A}\left( \boldsymbol{r}_{d},\boldsymbol{r}%
_{t};\omega\right) =g_{s\sigma}^{R\ast}\left( \boldsymbol{r}_{t},\boldsymbol{%
r}_{d};\omega\right) $ is the corresponding advanced Green's function.

The physical meaning of these three terms is transparent. The incoming part $%
A_{\sigma}^{\mathrm{inc}}$ gives the smooth background, since it carries no
resonant dependence on the impurity detuning. The scattered part $A_{\sigma
}^{\mathrm{sc}}$ gives a Lorentzian-type resonant contribution, because the
resonant impurity Green's function enters only through its modulus squared.
Finally, the interference term $A_{\sigma}^{\mathrm{int}}$ mixes the
incoming and scattered components and therefore produces the asymmetric Fano
contribution. In this way, the Lippmann--Schwinger equation makes explicit
that the local spectral-function asymmetry originates from the interference
between the incoming and scattered components in the substrate-projected
scattering state.

However, in many studies that focus mainly on the pole structure of the
Green's function, the line-shape asymmetry is easily overlooked. This is
especially the case when a wide-band approximation is applied and the real
part of $g_{s\sigma}^{R}$ is neglected, so that the corresponding level
shift is set to $\Lambda\rightarrow0$. In this limit, the Fano-type
structure in the $\lambda\rightarrow0$ limit of Eq.~(\ref%
{eq:local_fano_general}) reduces to a purely Lorentzian form: 
\begin{align}
& 1 -\func{Im}\!\bigg[ \frac{ \left[ \bar{g}_{s\sigma}^{R} (\boldsymbol{r}%
_{d},\boldsymbol{r}_{t}) \right] ^{2} }{ \Delta_{\sigma }/\Gamma+i } \bigg] 
\notag \\
& \quad\rightarrow1 + \left[ \func{Im} \bar{g}_{s\sigma}^{R} (\boldsymbol{r}%
_{d},\boldsymbol{r}_{t}) \right] ^{2} \func{Im}\!\bigg( \frac{1}{%
\Delta_{\sigma}/\Gamma+i} \bigg)  \notag \\
& \quad= 1 - \frac{ \left[ \func{Im} \bar{g}_{s\sigma}^{R} (\boldsymbol{r}%
_{d},\boldsymbol{r}_{t}) \right] ^{2} }{ (\Delta_{\sigma }/\Gamma)^{2}+1 }.
\end{align}

This result indicates that, in the absence of a direct tip-impurity
tunneling channel, the Fano-like asymmetry discussed here is primarily a
phase-sensitive effect of the substrate propagation amplitude, rather than a
consequence of the competition between a direct tunneling path and an
indirect tunneling path.

\section{Relation to the potential-scattering model}

\label{app:scatterer_mapping}

This appendix clarifies the relation between the present resonant-impurity
model in the $\lambda=0$ limit and the potential-scattering model. For
simplicity, we focus on the isotropic case with $\kappa_{\sigma}=1$ ($%
\mathcal{J}=0$) and neglect the Zeeman splittings. The spin indices in
quantities such as $\Delta_{\sigma}$ and $\bar{g}_{s\sigma}^{R}$ can then be
omitted, since the two spin channels become degenerate. Likewise, quantities
such as $A$ and $\bar{g}_{s}^{R}$ depend only on the tip-impurity distance $%
d $, because the spatial anisotropy is absent.

In many treatments, the impurity is modeled as a $\delta$-type scattering
potential. In such a potential-scattering description, no additional
impurity orbital is introduced, and the Hilbert space remains that of the
substrate. The full substrate Green's function can then be written as \cite%
{rusin2018theory} 
\begin{align}
& G_{s}^{R}\left( \boldsymbol{r},\boldsymbol{r}^{\prime};E_{F}\right)
=g_{s}^{R}\left( \boldsymbol{r},\boldsymbol{r}^{\prime};E_{F}\right)  \notag
\\
& +g_{s}^{R}\left( \boldsymbol{r},\boldsymbol{r}_{d};E_{F}\right) T\left(
E_{F}\right) g_{s}^{R}\left( \boldsymbol{r}_{d},\boldsymbol{r}^{\prime
};E_{F}\right) .
\end{align}
Here, 
\begin{equation}
T\left( E_{F}\right) =\frac{V_{r}}{1-V_{r}g^{(\mathrm{reg})}}
\label{eq:t_matrix_point_scatterer}
\end{equation}
is the corresponding $T$ matrix, where $V_{r}$ denotes the regularized
strength of the scattering potential. The same regularized local substrate
Green's function $g^{(\mathrm{reg})}$ also enters the resonant-impurity
self-energy through 
\begin{equation}
\left\vert \mathcal{A}_{ai}\right\vert ^{2}g^{(\mathrm{reg}
)}=\Lambda-i\Gamma,
\end{equation}
as discussed in Appendix~\ref{app:impurity_renormalization}.

In the potential-scattering model, the far-field elastic scattering
amplitude of quasiparticles at the Fermi energy is governed by the $T(E_{F})$%
. Thus, for a fixed substrate Green's function, the scattering strength is
controlled by the effective potential strength $V_{r}$. In the
weak-potential limit $\left\vert V_{r}\right\vert \rightarrow0$, or
equivalently in the Born regime $\left\vert V_{r}g^{(\mathrm{reg}%
)}\right\vert \ll1$ for finite $g^{(\mathrm{reg})}$, the $T$ matrix reduces
to 
\begin{equation}
T\left( E_{F}\right) = V_{r}\sum_{n=0}^{\infty}\left[ V_{r}g^{(\mathrm{reg}%
)} \right] ^{n} \simeq V_{r}.
\end{equation}
Thus, in the limit $\left\vert V_{r}\right\vert \rightarrow0$, the impurity
is effectively turned off. In the opposite limit $\left\vert
V_{r}\right\vert \rightarrow\infty$, the scattering enters the unitary
regime, where the $T$ matrix saturates as 
\begin{equation}
\lim_{\left\vert V_{r}\right\vert \rightarrow\infty} T\left( E_{F}\right) =
- \frac{1}{g^{(\mathrm{reg})}} .
\end{equation}

By contrast, in the present model, the physical picture is different. The
impurity has an explicit internal degree of freedom, and the corresponding $%
T $ matrix is given by 
\begin{equation}
T^{R}\left( E_{F}\right) =\left\vert \mathcal{A}_{ai}\right\vert
^{2}G_{d,d}^{R}\left( E_{F}\right) =\frac{\left\vert \mathcal{A}
_{ai}\right\vert ^{2}}{\Delta^{+}-\left\vert \mathcal{A}_{ai}\right\vert
^{2}g^{(\mathrm{reg})}},  \label{eq:t_matrix_resonant_impurity}
\end{equation}
where $\Delta^{+}=\Delta+i\eta$ with $\eta\rightarrow0^{+}$. In the main
text, this infinitesimal retarded prescription is omitted when $\left\vert 
\mathcal{A}_{ai}\right\vert $ is finite, since the retarded self-energy $%
\left\vert \mathcal{A}_{ai}\right\vert ^{2}g^{(\mathrm{reg})}$ already
provides a finite broadening. Here it is kept explicitly only to make the
weak-coupling limit $\left\vert \mathcal{A}_{ai}\right\vert \rightarrow0$
well defined.

Equation~(\ref{eq:t_matrix_resonant_impurity}) shows that the scattering
strength is governed not only by the impurity-substrate coupling strength $%
\left\vert \mathcal{A}_{ai}\right\vert$, but also by the detuning of the
impurity level. Indeed, by comparing Eq.~(\ref{eq:t_matrix_point_scatterer})
with Eq.~(\ref{eq:t_matrix_resonant_impurity}), one finds the mathematical
equivalence 
\begin{equation}
V_{r}(E_{F})=\frac{\left\vert \mathcal{A}_{ai}\right\vert ^{2}}{\Delta^{+}}.
\end{equation}
Thus, after eliminating the explicit impurity degree of freedom, the
resonant impurity model can be interpreted as an energy-dependent
regularized $\delta$-type scattering potential. In this representation, the
impurity is effectively turned off either when $\left\vert \mathcal{A}%
_{ai}\right\vert ^{2}\rightarrow0$ at fixed finite $\Delta$, or when $%
\left\vert \Delta\right\vert \rightarrow \infty$ at fixed finite $\left\vert 
\mathcal{A}_{ai}\right\vert$. Conversely, the scattering approaches the
unitary regime when this ratio becomes large, either by increasing $%
\left\vert \mathcal{A}_{ai}\right\vert ^{2}$ at fixed finite $\Delta$, or by
tuning $\Delta\rightarrow0$ at fixed finite $\left\vert \mathcal{A}%
_{ai}\right\vert$.

\begin{figure}[ptb]
\centering
\includegraphics[width=0.7\linewidth]{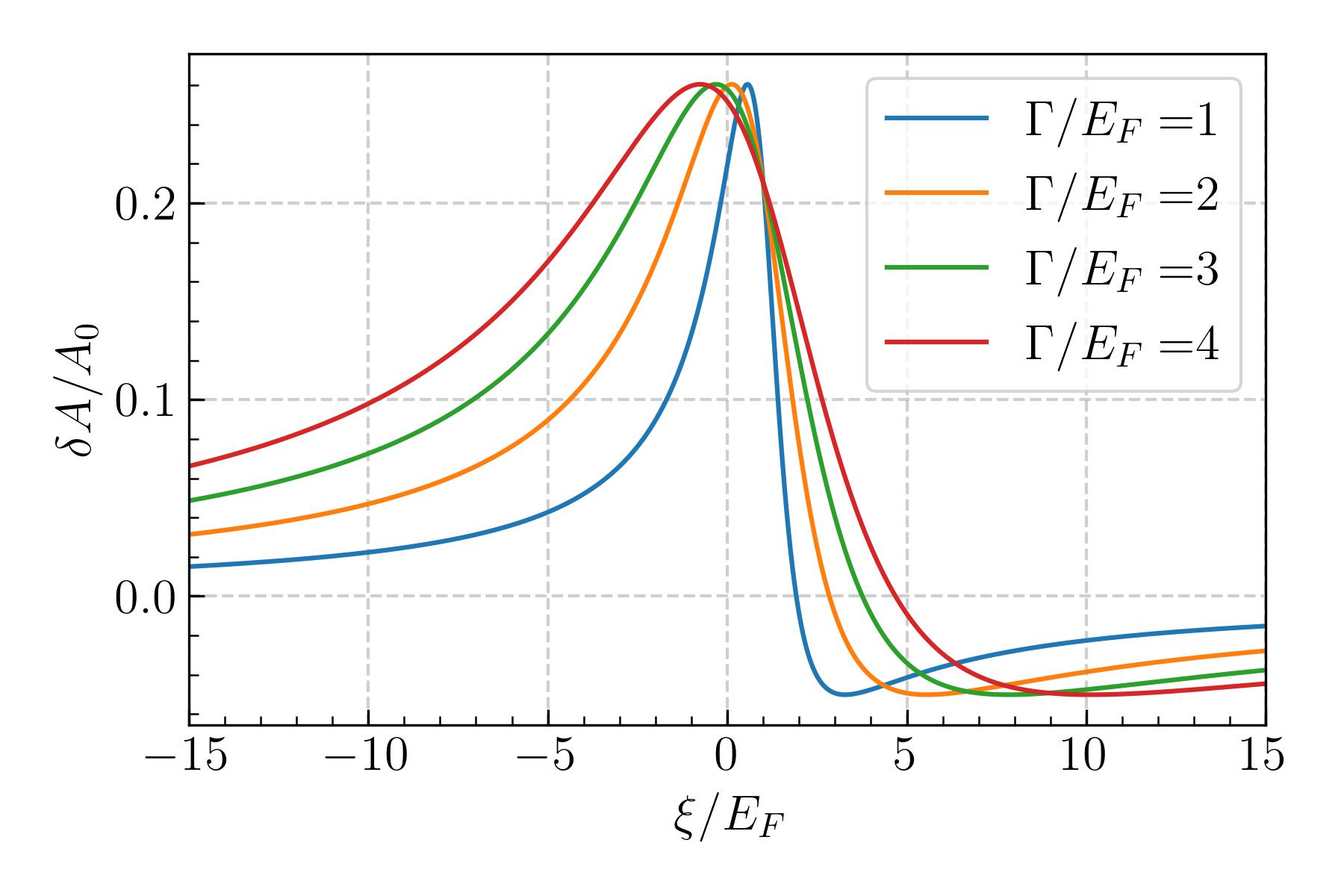}
\caption{Impurity-induced change in the local spectral function, $\protect%
\delta A$, as a function of $\Gamma/E_{F}$, which is proportional to $| 
\mathcal{A}_{ai}|^{2}$. The tip-impurity separation is fixed at $k_{F}d=2$.}
\label{fig:figs2}
\end{figure}

The local spectral-function response provides a direct way to visualize this
distinction. When it is plotted as a function of the bare detuning $\Delta$,
the linewidth and the apparent profile are controlled by $\left\vert 
\mathcal{A}_{ai}\right\vert ^{2}$, since both the energy shift $\Lambda$ and
the broadening factor $\Gamma$ are proportional to $\left\vert \mathcal{A}%
_{ai}\right\vert ^{2}$ (see Fig.~\ref{fig:figs2}). By contrast, when the
detuning axis is rescaled by $\Gamma$, or equivalently by using $(\Delta
-\Lambda)/\Gamma$ as the dimensionless detuning, the local spectral-function
curves collapse onto the same normalized line shape.

\begin{figure*}[ptb]
\centering
\includegraphics[width=1\linewidth]{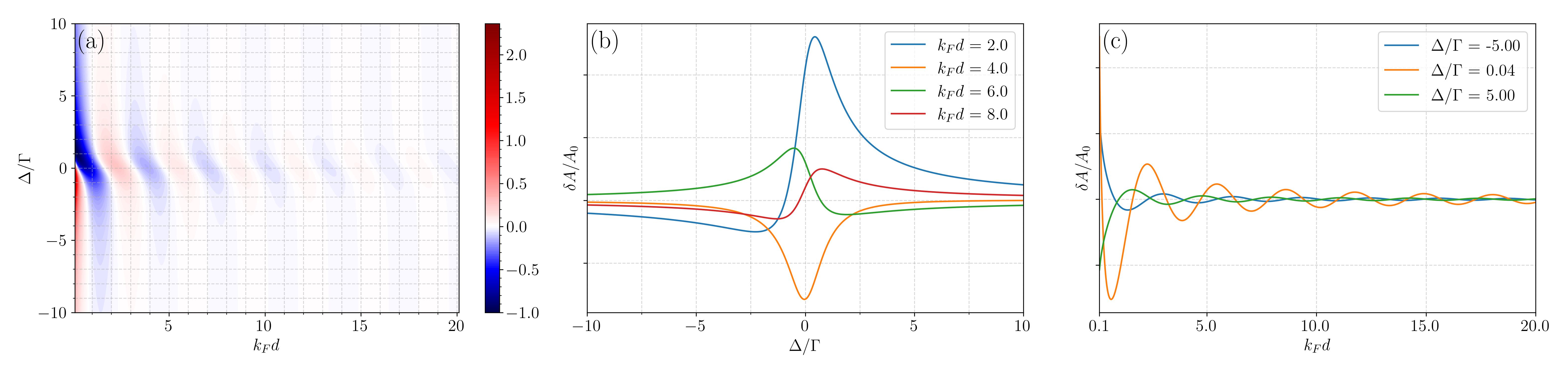}
\caption{(a) Local spectral function variation $\protect\delta A$ as a
function of $\Delta/\Gamma$ and $k_{F}d$ in the spin-degenerate case. (b),
(c) Line cuts as functions of $\Delta/\Gamma$ at fixed $k_{F}d$ and as
functions of $k_{F}d$ at fixed $\Delta/\Gamma$, respectively.}
\label{fig:supp_isotropic_fano}
\end{figure*}

Keeping the impurity level explicit also provides a more physical
interpretation of the local spectral-function evolution than directly tuning 
$V_{r}$ in the potential-scattering description. In the potential-scattering
description, the two unitary limits $V_{r}\rightarrow+\infty$ and $%
V_{r}\rightarrow-\infty$ correspond formally to infinitely strong repulsive
and attractive point potentials. In the resonant-impurity description,
however, they simply correspond to approaching the impurity-level alignment
from the positive- and negative-detuning sides, $\Delta\rightarrow0^{+}$ and 
$\Delta\rightarrow0^{-}$, respectively. In the weak-scattering regime,
namely when the impurity level is deeply off-resonant, the sign of the
response can be understood from virtual hopping between the substrate states
near the Fermi energy and the impurity level. If the impurity level lies
below the Fermi energy, $\xi<E_{F}$, then $\Delta>0$ and the virtual
intermediate state is lower in energy than the substrate state at $E_{F}$.
The associated second-order energy shift is therefore positive, which is
equivalent to a repulsive local potential, $V_{r}>0$. This repulsive
potential suppresses the substrate local spectral function in the vicinity
of the impurity. Conversely, if the impurity level lies above the Fermi
energy, $\xi>E_{F}$, then $\Delta<0$ and the virtual intermediate state is
higher in energy. The corresponding second-order energy shift is negative,
which is equivalent to an attractive local potential, $V_{r}<0$, and
therefore enhances the local spectral function near the impurity site [see
Fig.~\ref{fig:supp_isotropic_fano}(c)].

\section{Parameter estimates for the altermagnetic substrate}

\label{app:parameter_estimates}

Rather than fitting a specific compound, we adopt a phenomenological
parameter window guided by representative altermagnets and
altermagnet-related systems, including CrSb, KV$_{2}$Se$_{2}$O, Rb$%
_{1-\delta}$V$_{2}$Te$_{2}$O, Mn$_{5}$Si$_{3}$, and MnTe \cite%
{Yang2025CrSb3D,Reimers2024CrSbThinFilm,Terashima2026CrSbQO,
Jiang2025KV2Se2O,Zhang2025RbV2Te2O,Reichlova2024Mn5Si3AHE,
Badura2025Mn5Si3ANE,Lee2024MnTe}.

We first estimate the energy scale of a conventional metallic pocket.
Quantum-oscillation measurements in CrSb suggest relatively light effective
masses, with reported values of order $m^{\ast }/m_{e}\sim 0.24$--$1.23$ 
\cite{Terashima2026CrSbQO}, where $m_{e}$ is the bare electron mass. In the
following estimates we therefore focus on the low-mass regime and use $%
m^{\ast }=0.5m_{e}$ as a representative value. For conventional metallic
pockets, reported Fermi-surface contours and quantum-oscillation frequencies
typically correspond to Fermi wave vectors of order $k_{F}\sim 0.2$--$0.7~%
\mathrm{\mathring{A}}^{-1}$. Within the effective single-band parabolic
approximation, the Fermi energy measured from the bottom of the effective
band is $E_{F}=\hbar ^{2}k_{F}^{2}/2m^{\ast }$. Thus, for metallic pockets
with the above $k_{F}$ scale, one naturally obtains Fermi energies in the
range of several tenths of an electron volt or larger. In the main text, we
regard $E_{F}\sim 0.2$--$0.8~\mathrm{eV}$ as a representative
metallic-pocket energy scale.

We next estimate the dimensionless altermagnetic parameter. In the continuum
model used in the main text, the $d$-wave altermagnetic splitting is
described by $\varepsilon _{\boldsymbol{k}\sigma }=\hbar ^{2}\left(
k_{x}^{2}+k_{y}^{2}+2\sigma \mathcal{J}k_{x}k_{y}\right) /2m^{\ast }$, so
that the local spin splitting on a Fermi contour is of order $\Delta _{%
\mathrm{AM}}^{\mathrm{local}}(k_{F})\sim 2\mathcal{J}E_{F}$, up to an
angular form factor of order unity. Spectroscopic measurements have reported
sizable altermagnetic band splittings near the Fermi level, reaching about $%
1.0~\mathrm{eV}$ in CrSb, up to $1.6~\mathrm{eV}$ in KV$_{2}$Se$_{2}$O, and
about $0.37~\mathrm{eV}$ in MnTe \cite%
{Yang2025CrSb3D,Jiang2025KV2Se2O,Lee2024MnTe}. Since the present theory is a
local single-pocket continuum description, we do not directly use these
global maximal splittings. Instead, we adopt a local splitting scale on the
reference Fermi contour, $\Delta _{\mathrm{AM}}^{\mathrm{local}}(k_{F})\sim
0.05$--$0.6~\mathrm{eV}$. Combining this scale with the representative
metallic-pocket estimate of $E_{F}$ gives $\mathcal{J}\sim \Delta _{\mathrm{%
AM}}^{\mathrm{local}}(k_{F})/2E_{F}\sim 0.1$--$0.6$ as a reasonable
phenomenological range. Thus, the value $\mathcal{J}=0.4$ used in the main
text lies within a representative parameter window and keeps the effective
dispersion in the stable regime $\mathcal{J}<1$.

For the numerical calculations in the main text, however, we focus on a
shallow pocket regime rather than on a conventional large metallic pocket.
Here $E_{F}$ should be understood as the chemical potential measured from
the bottom of the effective pocket being probed. This choice is motivated by
two considerations. First, a smaller $k_{F}$ enlarges the real-space length
scale of the oscillatory or nodal local spectral function patterns, making
them more favorable for STM imaging. Second, for a conventional metallic
pocket with $E_{F}\sim 0.2\text{--}0.8~\mathrm{eV}$, a magnetic field of a
few tesla would lead to a very large Landau-level filling factor, $\zeta
=E_{F}/\hbar \omega _{L}$, which is not the moderate filling regime
considered in the main text. Thus, the values of $\zeta $ used below should
be interpreted as describing a shallow effective pocket. Such a regime can
be reached when the chemical potential lies close to the bottom of the
relevant band, or when the local tunneling signal is dominated by a
low-density or surface-related pocket.

Taking a moderate magnetic field $B=5~\mathrm{T}$ as an example, and using $%
\hbar \omega _{L}=\sqrt{1-\mathcal{J}^{2}}\cdot m_{e}/m^{\ast }(0.1158~%
\mathrm{meV/T})B$ with $m^{\ast }=0.5m_{e}$ and $\mathcal{J}=0.4$, we obtain 
$\hbar \omega _{L}\simeq 1.06~\mathrm{meV}$. For the representative filling
factor used in the main text, $\zeta =5$, the corresponding Fermi energy is $%
E_{F}=\zeta \hbar \omega _{L}\simeq 5.3~\mathrm{meV}$. The associated Fermi
wave vector is $k_{F}=\sqrt{2m^{\ast }E_{F}/\hbar ^{2}}\simeq 0.026~\mathrm{%
\mathring{A}}^{-1}$. This value is much smaller than that of a conventional
large metallic pocket. The corresponding real-space length scale is $%
k_{F}^{-1}\simeq 3.8~\mathrm{nm}$, which is well within typical
scanning-probe length scales.

We also discuss the phenomenological broadening. In the main calculations we
take $\eta =10^{-3}E_{F}$. For the representative shallow pocket parameters
above, this gives $\eta \simeq 5.3~\mu \mathrm{eV}$. Thus, $\hbar \omega
_{L}/\eta \simeq 1.06~\mathrm{meV}/5.3~\mu \mathrm{eV}\simeq 200$.
Equivalently, since $\eta =10^{-3}E_{F}=10^{-3}\zeta \hbar \omega _{L}$, one
has $\hbar \omega _{L}/\eta =10^{3}/\zeta $. For $\zeta =5$, the
Landau-level spacing is therefore two hundred times larger than the
phenomenological broadening. The calculations are therefore performed in the
resolved-Landau-level regime, $\hbar \omega _{L}\gg \eta $. In realistic
systems, disorder, temperature, and instrumental resolution contribute to an
effective broadening $\eta _{\mathrm{eff}}$; the essential requirement for
observing the Landau-quantized local spectral function structure is $\eta _{%
\mathrm{eff}}\ll \hbar \omega _{L}.$

\bibliographystyle{apsrev}
\bibliography{HY_reference1}

\end{document}